\documentclass[a4paper,10pt]{article} 

\usepackage{amsthm}
\usepackage{amsfonts}
\usepackage{amssymb}
\usepackage{amsthm}

\usepackage{youngtab}
\usepackage[T1]{fontenc}

\newcommand{\DAl}{\raise -0.4mm \hbox{\Large$\Box$}}

\def\Proof{{\em Proof:\hspace{2mm}}}

\newtheorem{theorem}{Theorem}
\newtheorem{prop}{Proposition}
\newtheorem{remark}{Remark}
\newtheorem{definition}{Definition}

\begin{document}

\title{Spinor calculus on 5-dimensional spacetimes}

\author{Alfonso Garc\'{\i}a-Parrado G\'omez-Lobo \footnote{Present address: Centro de Matem\'atica, Universidade do Minho 4710-057 Braga, Portugal}
\thanks{E-mail address:
{\tt alfonso@cage.ugent.be}}\\
Ghent University, Department of Mathematical Analysis.\\ 
Galglaan 2, 9000 Ghent, Belgium.
\\
\\
Jos\'e M. Mart\'{\i}n-Garc\'{\i}a
\thanks{E-mail address:
{\tt Jose.Martin-Garcia@obspm.fr}}\\
Laboratoire Univers et Th\'eories, Observatoire de Paris, CNRS, \\
Univ. Paris Diderot, 5 place Jules Janssen, 92190 Meudon, France,
 and \\
Institut d'Astrophysique de Paris, Univ. Pierre et Marie Curie, CNRS \\
98bis boulevard Arago, 75014 Paris, France
}

\maketitle


\begin{abstract}
Penrose's spinor calculus of 4-dimensional Lorentzian geometry is extended to the case of 5-dimensional Lorentzian geometry. Such fruitful ideas in Penrose's spinor calculus as the spin covariant derivative, the curvature spinors or the definition of the spin coefficients on a spin frame can be carried over to the spinor calculus in 5-dimensional Lorentzian geometry. 
The algebraic and differential properties of the curvature spinors are studied in detail and as an application we extend the well-known 4-dimensional Newman-Penrose formalism to a 5-dimensional spacetime.
\end{abstract}

PACS: 02.10.Xm, 04.20.Cv, 04.20.Gz, 02.70.Wz, 04.50.-h, 02.40.Hw.

\maketitle

\section{\label{sec:int}Introduction}
In recent years high progress has been made towards the understanding of General Relativity when the number of spacetime dimensions is five. As a result it is now known that some classical results of 4-dimensional General Relativity do not translate to a 5-dimensional spacetime (or at least the translation is not  straightforward). Perhaps the best known example is supplied by the uniqueness theorems of stationary vacuum black holes which, in four dimensions, put severe constraints on the topology of the event horizon and the nature of the spacetime when certain conditions are met (an up-to-date review of this subject is \cite{CHRUSCIEL-COSTA}). In five dimensions these theorems are no longer true as it was first shown by means of a counterexample in \cite{EMPARAN-REALL}. The counterexample was an exact solution of the Einstein vacuum field equations in five dimensions which is stationary, asymptotically flat and has a connected, non-degenerate event horizon (it is thus a black hole), yet the topology of the event horizon is $S^1\times S^2\times \mathbb R$ (a black ring). Since the solution is asymptotically flat one can define its mass and its angular momentum and check that there is a range of these variables which fall within the ranges of the mass and the angular momentum of the Myers-Perry solution in dimension five. In this way one concludes that in five dimensions it is possible to have two different (non-isometric) vacuum, stationary, asymptotically flat black holes with the same mass and angular momentum and both having a non-degenerate connected event horizon. 

The discovery of the black ring fostered the investigation of exact solutions in five and higher dimensions (see \cite{HIGHER-D-REVIEW} for a thorough review of the research conducted in this direction). In this framework it is useful to have generalisations to higher dimensions of the tools which have been successfully employed in four dimensions to find and classify exact solutions. Two of these tools are the {\em Newman-Penrose formalism} and the {\em Petrov classification}. For a generic spacetime dimension, which we may call $N$, these tools have been recently generalised by a number of authors.  The introduction of the {\em alignment theory} in \cite{ALIGNED-THEORY} made it possible to put forward a scheme to classify the Weyl tensor (and in fact any tensor) in a Lorentzian vector space of arbitrary dimension 
(this is reviewed in \cite{COLEY-REVIEW}). A frame formalism in which the Ricci and Bianchi identities are studied for a spacetime of dimension $N$ has been developed in \cite{RICCI-N} and \cite{BIANCHI-N} respectively.
   
While the afore-mentioned generalisations are useful when working in generic dimension $N$ one hopes that in the particular case of $N=5$ it should be possible to adopt the same procedure which is followed in dimension four to introduce the Newman-Penrose formalism and the Petrov classification. This consists in regarding these as natural applications of the {\em spinor calculus} introduced by Penrose in \cite{PENROSE-SPINOR} and therefore this approach requires the development of the spinor calculus in a 5-dimensional Lorentzian manifold. To present a detailed description of this spinor calculus is one of the main aims of this paper. Using this spinor formalism we extend the Newman-Penrose formalism to a 5-dimensional spacetime. We stress that the formalism introduced here is truly an ``extension'' of the Newman-Penrose formalism used in four dimensions because it contains the same variables (spin coefficients, curvature scalars, etc) as in four dimensions plus some additional quantities which are specific to a 5-dimensional spacetime. As can be expected, the full set of Newman-Penrose equations in a 5-dimensional spacetime is far more involved than in four dimensions but this does not mean that the formalism will be less useful in certain particular cases as we hope to illustrate in \cite{AGPLODE}. One can also develop an invariant classification of the Weyl spinor associated to the 5-dimensional Weyl tensor \cite{SMET1}.

It is well-known that spinors can be introduced in a spacetime of arbitrary dimension ---in fact the signature of the metric tensor need not be Lorentzian. The generic procedure to accomplish this is also well-known (see appendix A of \cite{PENROSERINDLER2} for an account of it) but if we are interested in the particular case of a 5-dimensional spacetime it is worth spelling out the whole procedure in detail for this particular case, specially if we are to focus on specific applications 
such as those mentioned above. 
For a 5-dimensional spacetime, the spin space is a 4-dimensional complex vector space endowed with an antisymmetric tensor which plays the role of a metric tensor (symplectic structure).
Also we show how one can develop a {\em calculus} using these spinors and explain how to introduce the idea of {\em a spin covariant derivative}. When the spin covariant derivative is compatible with the spacetime metric and the symplectic structure then we prove that such a spin covariant derivative is unique. In this case one can define the {\em curvature spinors} (Ricci spinor and Weyl spinor) much in the same way as it is done in the spinor calculus of a 4-dimensional Lorentzian manifold. The algebraic and differential properties of these spinors are analysed and it is found that the Weyl spinor is a rank-4 totally symmetric spinor (this property was already pointed out in \cite{SMET1}) and that the Ricci spinor is a rank-4 {\em Riemann-like tensor}, by which we mean that it has the same symmetries as a 4-dimensional Riemann tensor (we regard the {\em cyclic property} of the Riemann tensor as one of its symmetries).

Working in dimension five leads to the manipulation of tensors with
large numbers of components, already with only three or four indices.
This, together with the presence of various types of symmetries
(including an antisymmetric metric), makes it convenient to use
specialised tools for Tensor Computer Algebra. We have used the
system {\em xAct} \cite{xAct}, based on {\em Mathematica}, and
developed by one of us (JMM). {\em xAct} can handle both abstract and component
expressions with arbitrary permutation symmetries, by means
of efficient techniques of computational group theory \cite{xPerm},
and by using systematically Penrose's abstract index notations, as
we shall do in this article.
We have recently extended it to perform spinor calculus in 4-dimensional spacetimes
\cite{Spinors}, and the 5-dimensional spinor calculus studied in the present paper will be also included in a near future.

The structure of this paper is as follows: in section \ref{sec:spin-5d} we recall how spinors are constructed in a 5-dimensional vector space possessing a Lorentzian metric and discuss some basic properties of the {\em spin space}. Section \ref{sec:spin-structure} deals with the concept of {\em spin structure} on a 5-dimensional Lorentzian manifold. One then can define a covariant derivative which is compatible with the spin structure (spin covariant derivative). We prove here that a spin covariant derivative which is also compatible with the symplectic structure of the spin bundle is unique and extends the Levi-Civita connection to the spin bundle. To carry out the proof we compute the connection coefficients of such a spin covariant derivative in a {\em spin tetrad} and a {\em semi-null pentad} which enables us to extend the Newman-Penrose spin coefficients to the 5-dimensional case.  
In section \ref{section:curvature-spinors}
we use this covariant derivative to find the spinors representing the traceless Ricci tensor and the Weyl tensor (curvature spinors) and study the algebraic and differential properties of these spinors.
The main application of previous results is shown in section \ref{sec:newman-penrose}
where we explain how one can extend the Newman-Penrose formalism to a 5-dimensional spacetime. 
Some further applications are indicated in section \ref{sec:conclusions}.

\section{The spin structure on a 5 dimensional Lorentzian vector space}
\label{sec:spin-5d}
Let $\mathbf L$ be a 5-dimensional real vector space endowed with a real scalar product $g(\ ,\ )$ of Lorentzian signature (signature convention $(+,-,-,-,-)$) and let $\mathbf S$ be a complex vector space whose dimension is for the moment left unspecified (complex conjugate of scalars will be denoted by an overbar). Using the vector space $\mathbf L$ and its dual ${\mathbf L}^*$ as the starting point one builds a tensor algebra in the standard fashion. Similarly a tensor algebra is built from $\mathbf S$ and its dual ${\mathbf S}^*$. We denote these algebras by $\mathfrak{T}(\mathbf{L})$ and $\mathfrak{T}(\mathbf{S})$ respectively \footnote{Strictly speaking only the algebras $\mathfrak{T}^r_s(\mathbf L)$ of tensors $r$-contravariant $s$-covariant can be defined (and the same applies to $\mathfrak{T}^{r}_{s}(\mathbf S)$). To lessen the notation we will suppress the labels $r$, $s$ in the notation and they will only be made explicit when confusion may arise.}. 
In this work abstract indices will be used throughout to denote tensorial quantities: in this way small Latin indices $a,b,\dots$ will denote abstract indices on elements of $\mathfrak{T}(\mathbf{L})$ and capital Latin indices $A,B,\dots$ will be used for abstract indices of elements in $\mathfrak{T}(\mathbf{S})$. The tensor algebra $\mathfrak{T}(\mathbf{S})$ will be referred to as the {\em spin algebra} and its elements will be called spinors. One can also build tensor algebras by taking tensor products of elements in $\mathfrak{T}(\mathbf L)$ and elements in $\mathfrak{T}(\mathbf S)$. Quantities in these tensor algebras will be referred to as {\em mixed tensors} and they will carry abstract indices of both types. The algebras $\mathfrak{T}(\mathbf L)$ and $\mathfrak{T}(\mathbf S)$ shall be regarded as complex vector spaces.

We introduce now a mixed tensor $\gamma_{aB}^{\phantom{aB}C}$ which, by definition, fulfils the following algebraic property

\begin{equation}
\gamma_{aA}^{\phantom{aA}B} \gamma_{bB}^{\phantom{bB}C}
+
\gamma_{bA}^{\phantom{bA}B} \gamma_{aB}^{\phantom{aB}C}
=- \delta_{A}^{\phantom{A}C}g_{ab}, 
\label{eq:clifford}
\end{equation}
where $\delta_{A}^{\phantom{A}C}$ is the identity tensor (also known as the {\em Kronecker delta})
on the vector space $\mathbf S$. Note that we use a {\em staggerred}

This relation means that $\gamma_{aB}^{\phantom{aB}C}$ can be regarded as belonging to a representation on the vector space $\mathbf S$ of the 
{\em Clifford algebra} $Cl(\mathbf L,g)$. 
If we demand that this representation be {\em irreducible} then  
it can be shown that extra structures can be added to the vector space ${\mathbf S}$. 
First of all we note that the quantity $\gamma_{aA}^{\phantom{aA}A}$ must vanish, for otherwise the 1-form $\gamma_{aA}^{\phantom{aA}A}$ would be invariant under the action of any endomorphism of $\mathbf L$ keeping $g_{ab}$ invariant (orthogonal group), and this can only happen for scalars and $5$-forms \cite{PENROSERINDLER2}. Another consequence of 
$\gamma_{aB}^{\phantom{aB}C}$ belonging to an irreducible representation of the Clifford algebra is shown in the next result. 

\begin{theorem}
If the quantity $\gamma_{aB}^{\phantom{aB}C}$ belongs to an irreducible representation of $Cl(\mathbf L,g)$, then the dimension of ${\mathbf S}$ is 4 and there exist two antisymmetric spinors $\epsilon_{AB}$, $\widehat\epsilon^{AB}$, unique up to a constant, such that
\begin{equation}
\epsilon_{AB}\widehat\epsilon^{CB}=\delta^{\phantom{A}C}_{A}.
\label{eq:epsinverse}
\end{equation}
Furthermore, these antisymmetric spinors fulfil the following algebraic property
\begin{equation}
\gamma{}_{a}{}_{D}{}^{A}\gamma{}^{a}{}_{C}{}^{B}=
\frac{1}{2} \delta{}_{D}{}^{A} \delta{}_{C}{}^{B}
-\delta {}_{C}{}^{A}\delta {}_{D}{}^{B}+\epsilon{}_{C}{}_{D} 
\widehat{\epsilon }{}^{A}{}^{B}. 
\label{eq:gamma-times-gamma}
\end{equation}
\label{theorem:spin-space} 
\end{theorem}

\noindent
\Proof Equation (\ref{eq:gamma-times-gamma}) is just equation (B.29) of the appendix of \cite{PENROSERINDLER2} particularised for the case in which ${\mathbf L}$ is 5-dimensional. Equation (B.41.b) of that appendix shows that the quantity 
${\mathbf E}^-$ of (B.29) reduces to the last term of (\ref{eq:gamma-times-gamma}). 
\qed

The results of previous theorem allow us to introduce the concept of {\em spin structure} on the vector space ${\mathbf L}$.

\begin{definition}
Under the conditions stated in theorem \ref{theorem:spin-space}  we will refer to   
$\gamma_{aA}^{\phantom{aA}B}$ as a spin structure on ${\mathbf L}$. The complex vector space ${\mathbf S}$ is then called the spin space of the spin structure.  
\label{def:spin-structure-L}
\end{definition}
\noindent
The spinors $\epsilon_{AB}$ and $\widehat\epsilon^{AB}$ can be regarded as a metric tensor and its inverse in the vector space $\mathbf S$ and therefore they can be used to raise and lower spinorial indices. The metric tensor $\epsilon_{AB}$ is a {\em symplectic metric} and therefore some care is required when introducing the conventions for the raising and lowering of indices with $\epsilon_{AB}$ and $\widehat\epsilon^{AB}$ (see appendix \ref{appendix:symplectic} for a review of this issue). Our conventions for these o\-pe\-rations are
$$
\xi^A\epsilon_{AB}=\xi_B\;,\quad \xi^A=\widehat\epsilon^{AB}\xi_B.
$$  
In particular we can raise the indices of $\epsilon_{AB}$ getting 
$\epsilon^{AB}=\widehat\epsilon^{AB}$ and from now on only the symbol $\epsilon$ will be used for the symplectic metric and its inverse. Note also the property 
\begin{equation}
\delta^A_{\phantom{A}B}=-\delta_B^{\phantom{B}A}.
\label{eq:kronecker-delta}
\end{equation}
Here the quantity $\delta_B^{\phantom{B}A}$ is the {\em Kronecker delta} on 
$\mathbf S$ and $\delta^A_{\phantom{A}B}$ is a derived quantity obtained from it by the raising and lowering of indices. In particular this implies $\delta_A^{\phantom{A}A}=4$. Again see  appendix \ref{appendix:symplectic} for a further discussion about the properties and conventions related to a symplectic metric.
 
Using $\epsilon_{AB}$, $\epsilon^{AB}$, $g_{ab}$ and $g^{ab}$ we can raise and lower indices of mixed quantities. In particular, we can start from $\gamma_{aA}^{\phantom{aA}B}$ and obtain the quantity $\gamma^a_{\phantom{a}AB}$. Next result gathers a number of algebraic properties of $\gamma^a_{\phantom{a}AB}$ which are needed in this work. 

\begin{theorem}
The quantity $\gamma^a_{\phantom{a}AB}$ has the following algebraic properties
\begin{eqnarray}
&&\gamma_{a}^{\phantom{a}AB}\gamma_{bAB}= -2 g_{ab}\;,\label{eq:gamma2-contracted}\\
&&\gamma_{aCD}\gamma^{a}_{\phantom{a}AB}=\epsilon_{AD}\epsilon_{BC}-
\epsilon_{AC}\epsilon_{BD}+\frac{1}{2}\epsilon_{AB}\epsilon_{CD}\;,
\label{eq:gammasquare}\\
&&\gamma^a_{\phantom{a}[AB]}=\gamma^a_{\phantom{a}AB}\;,
\quad
\epsilon^{AB}\gamma^{a}_{\phantom{a}AB}=0.\label{eq:gammata}
\end{eqnarray}
\label{theorem:gamma-properties}
\end{theorem}

\noindent
\Proof Equation (\ref{eq:gamma2-contracted}) is a direct consequence from the trace of (\ref{eq:clifford}) while equation (\ref{eq:gammasquare}) comes from lowering all the indices in (\ref{eq:gamma-times-gamma}). Now, 
we multiply both sides of this last expression by $\gamma_{b}^{\phantom{b}AB}$, use 
(\ref{eq:gamma2-contracted})
and perform all the metric contractions. The result is
$$
\gamma_{bCD}=-\gamma_{bDC},
$$ 
which entails $\gamma^a_{\phantom{a}[AB]}=0$. Finally we note 
\begin{equation}
0=\gamma^{aA}_{\phantom{aA}A}=-\epsilon^{AB}\gamma^a_{\phantom{a}AB}.
\end{equation}
\qed

\noindent
For later applications we need to introduce another quantity related to $\gamma^a_{\phantom{a}AB}$. Its definition is 
\begin{equation}
G{}^{a}{}^{b}{}_{A}{}_{C}\equiv-\gamma {}^{a}{}_{(A}{}^{B} \
\gamma {}^{b}{}_{C)}{}_{B}.
\label{eq:G-definition}
\end{equation}
From this definition is clear that $G{}^{a}{}^{b}{}_{A}{}_{C}$ has the following symmetries
$$
G{}^{[a}{}^{b]}{}_{A}{}_{C}=G{}^{a}{}^{b}{}_{A}{}_{C}\;,\quad
G{}^{a}{}^{b}{}_{(A}{}_{C)}=G{}^{a}{}^{b}{}_{A}{}_{C}.
$$
Also, using the results of theorem \ref{theorem:gamma-properties} we deduce the following algebraic properties for $G^{ab}_{\phantom{ab}AB}$
\begin{eqnarray}
&& G{}_{a}{}_{b}{}_{C}{}_{D} G{}^{a}{}^{b}{}_{A}{}_{B}=
\epsilon{}_{A}{}_{D}\epsilon {}_{B}{}_{C} +
\epsilon {}_{A}{}_{C}\epsilon{}_{B}{}_{D}\;,\\
&& G_{ad}^{\phantom{ad}AB}G_{bcAB}=g{}_{a}{}_{b}g{}_{c}{}_{d}-g{}_{a}{}_{c} g{}_{b}{}_{d}.
\label{eq:Gsquares}
\end{eqnarray}
Finally we note that the product $\gamma {}_{a}{}_{A}{}^{B}\gamma {}_{b}{}_{C}{}_{B}$
can be written in the form
\begin{equation}
\gamma {}_{a}{}_{A}{}^{B}\gamma {}_{b}{}_{C}{}_{B}=-G{}_{a}{}_{b}{}_{A}{}_{C} +\frac{1}{2} g{}_{a}{}_{b}\epsilon {}_{A}{}_{C}.
\label{eq:productof2gammas}
\end{equation}

\subsection{Relation between spinors and tensors}
\label{subsection:spinors-tensors}
Elements of the tensor algebra $\mathfrak{T}(\mathbf{L})$ can be put into correspondence with spinors by means of $\gamma_{a}^{\phantom{a}AB}$. To see an example of how this works we consider a 1-rank vector $v^a$. Then its spinor counterpart is given by
$$
v^{AB}=\gamma_a^{\phantom{a}AB}v^a.
$$ 
In view of the properties presented in Theorem \ref{theorem:gamma-properties} it is immediate that the spinor $v^{AB}$ is antisymmetric and traceless. Reciprocally, any antisymmetric and traceless spinor $\xi^{AB}$ has a vector counterpart given by 
$\xi^{AB}\gamma^a_{\phantom{a}AB}$.

Previous example can be generalised for tensors and spinors of higher rank. 
The fundamental result is comprised in the following proposition

\begin{prop}
 Let $T_{A_1B_1\dots A_pB_p}\in\mathfrak{T}^0_{2p}({\mathbf S})$, $p\in\mathbb N$  
and suppose that
\begin{eqnarray}
 T_{A_1B_1\dots [A_jB_j]\dots A_pB_p}=T_{A_1B_1\dots A_jB_j\dots A_pB_p},
\label{eq:t2fold-a}\\
T^{\phantom{A_1B_1\dots A_{j-1}B_{j-1}}A_{j}}_{A_1B_1\dots 
A_{j-1}B_{j-1}\phantom{A_{j}} A_{j}A_{j+1}B_{j+1}\dots A_pB_p}=0,
\label{eq:t2fold-b}
\end{eqnarray}
for any $j=1,\dots p$. Then there is a unique 
tensor $T_{a_1\dots a_p}\in\mathfrak{T}^0_{p}(\mathbf{L})$ such that 
\begin{equation}
T_{a_1\dots a_p}\gamma^{a_1}_{\phantom{a_1}A_1B_1}\cdots\gamma^{a_p}_{\phantom{a_p}A_{p}B_{p}}=T_{A_1B_1\dots A_pB_p}. 
\label{eq:tensortospinor}
\end{equation}
The tensor $T_{a_1\dots a_p}$ is given by
\begin{equation}
T_{a_1\dots a_p}=\frac{1}{(-2)^p}
\gamma_{a_1}^{\phantom{a_1}A_1B_1}\cdots\gamma_{a_p}^{\phantom{a_p}A_pB_p} 
T_{A_1B_1\dots A_pB_p}.
\label{eq:tensor=spinor}
\end{equation}
\label{prop:antisymmetric-tracefree}
\end{prop}

\noindent
\Proof If $T_{A_1B_1\dots A_{p}B_{p}}$ is a given spinor fulfiling (\ref{eq:t2fold-a})-(\ref{eq:t2fold-b}) and we define a tensor $T_{a_1\dots a_p}$ through (\ref{eq:tensor=spinor}) then we have  
\begin{eqnarray*}
T_{a_1\dots a_p}\gamma^{a_1}_{\phantom{a_1}A_1B_1}\cdots\gamma^{a_p}_{\phantom{a_p}A_{p}B_{p}}=
\frac{1}{(-2)^p}
\gamma^{a_1}_{\phantom{a_1}A_1B_1}\gamma_{a_1}^{\phantom{a_1}C_1D_1}
\cdots\gamma^{a_p}_{\phantom{a_p}A_{p}B_{p}}\gamma_{a_p}^{\phantom{a_p}C_pD_p} 
T_{C_1D_1\dots C_pD_p}=\\
\left(
\frac{\delta {}_{A_1}{}^{C_1} \delta {}_{B_1}{}^{D_1}}{2}-
\frac{\delta {}_{A_1}{}^{D_1} \delta {}_{B_1}{}^{C_1}}{2} 
+\frac{\epsilon {}_{A_1}{}_{B_1}}{4} \
\epsilon{}^{D_1}{}^{C_1}\right)\cdots
\left(
\frac{\delta {}_{A_p}{}^{C_p} \delta {}_{B_p}{}^{D_p}}{2}-
\frac{\delta {}_{A_p}{}^{D_p} \delta {}_{B_p}{}^{C_p}}{2} 
+\frac{\epsilon {}_{A_p}{}_{B_p}}{4} \
\epsilon{}^{D_p}{}^{C_p}\right)
\times\\
T_{C_1D_1\dots C_pD_p}=T_{A_1B_1\dots A_pB_p},
\end{eqnarray*}
where (\ref{eq:gammasquare}) was used in the first step and  (\ref{eq:t2fold-a})-(\ref{eq:t2fold-b})  
in the last step. Suppose now that there is another tensor $\tilde{T}_{a_1\dots a_p}$ such that 
$$
\tilde T_{a_1\dots a_p}
\gamma^{a_1}_{\phantom{a_1}A_1B_1}\cdots\gamma^{a_p}_{\phantom{a_p}A_{p}B_{p}}
=T_{A_1B_1\dots A_{p}B_p}.
$$
Then if we multiply both sides of previous equation by 
$\gamma_{b_1}^{\phantom{a_1}A_1B_1}\cdots\gamma_{b_p}^{\phantom{a_p}A_{p}B_{p}}$ 
and use (\ref{eq:gamma2-contracted}) we obtain 
$$
(-2)^p\tilde T_{b_1\dots
b_p}=\gamma_{b_1}^{\phantom{b_1}A_1B_1}\cdots\gamma_{b_p}^{\phantom{b_p}A_pB_p} 
T_{A_1B_1\dots A_pB_p},
$$
which via equation (\ref{eq:tensor=spinor}) entails
$\tilde{T}_{a_1\dots a_p}=T_{a_1\dots a_p}$.\qed

Elements of $\mathfrak{T}^0_{2p}(\mathbf S)$ fulfiling (\ref{eq:t2fold-a})-(\ref{eq:t2fold-b}) form a subspace which is 
denoted by $\mathfrak{U}_{2p}(\mathbf S)$. Thus previous proposition asserts that $\mathfrak{U}_{2p}(\mathbf S)$ and $\mathfrak{T}^0_p(\mathbf L)$ are in fact isomorphic as vectors spaces. Similar considerations as before, lead us to the definition of $\mathfrak{U}^{2p}(\mathbf S)$.

It is possible to define a {\em unique} tensor starting from any spinor $\xi_{A_1B_1\dots A_pB_p}$ in $\mathfrak{T}^0_{2p}(\mathbf S)$ as follows. First  we introduce the {\em linear projector} ${\mathcal P}$ which projects the vector space $\mathfrak{T}^0_{2p}(\mathbf S)$ down to the subspace $\mathfrak{U}_{2p}(\mathbf S)$.
Then we apply proposition \ref{prop:antisymmetric-tracefree} to the spinor  ${\mathcal P}(\xi_{A_1B_1\dots A_pB_p})$. The tensor so obtained is called the {\em tensor counterpart} or {\em tensor equivalent} of $\xi_{A_1B_1\dots A_pB_p}$. It is clear that different spinors can have the same tensor counterpart.

\begin{prop}
 If $\xi^A$, $\eta^A$ are such that $\xi^A\eta_A=0$, then the spinor 
$\omega_{AB}\equiv\xi_{[A}\eta_{B]}$ 
defines a null $l^a$ vector by means of the relation
\begin{equation}
l^a=\gamma^a_{\phantom{a}AB}\omega^{AB} 
\label{eq:nullvector}
\end{equation}
\label{prop:nullvector}
\end{prop}

\noindent
\Proof We need to show that the vector $l^a$ defined by (\ref{eq:nullvector}) is null. Using (\ref{eq:gammasquare}) we get
$$
l^al_a=-2\omega_{AB}\omega^{AB}=-2\xi_{[A}\eta_{B]}\xi^{[A}\eta^{B]}=0.
$$
\qed
\begin{remark}\em
Unlike as in the case of the spinor algebra in 4-dimensional Lorentzian geometry, there is no converse to previous proposition. All what can be said is that if $l^a$ is null, then the spinor 
$\omega_{AB}\equiv \gamma_{aAB}l^a$ fulfils the property 
$$
\omega_{AB}\omega^{AB}=0,
$$
as is easily checked using (\ref{eq:gamma2-contracted}).
\end{remark}

\subsection{Spin tetrads and semi-null pentads}
\label{subsection:spin-pentad}
In this subsection we introduce a basis in $\mathbf S$, with elements
$o^A$, $\iota^A$, $\tilde{o}^A$, $\tilde{\iota}^A$, in which the
symplectic metric takes the canonical form
\begin{equation}
\epsilon_{AB}=2o_{[A}\iota_{B]}-2\tilde{o}_{[A}\tilde{\iota}_{B]}. 
\label{eq:epsilondyad}
\end{equation}
This entails
\begin{equation}
o^A\iota_A=-1=\tilde{o}^A\tilde{\iota}_A\;,\quad o^A\tilde{o}_A=
\iota^A\tilde{\iota}_A=o^A\tilde{\iota}_A=\iota^A\tilde{o}_A=0
\label{eq:dyadnormalization}
\end{equation}
The basis $\{o^A,\iota^A,\tilde{o}^A,\tilde{\iota}^A\}$ is the analog of the spin dyad 
which is used in the spinor calculus of 4-dimensional Lorentzian geometry. We will call a basis with these properties a {\em spin tetrad}. 
It is now clear that the spin space $\mathbf S$ can be written as the following direct sum
$$
{\mathbf S}={\mathbf S}_1\oplus{\mathbf S}_2\;,\quad {\mathbf S}_1=\mbox{span}\{o^A,\iota^A\}\;,\quad
{\mathbf S}_2=\mbox{span}\{\tilde{o}^A,\tilde{\iota}^A\}.
$$
Each of the spaces ${\mathbf S}_1$, ${\mathbf S}_2$ is isomorphic to the 2-dimensional spin space in which the spinors of 4-dimensional Lorentzian geometry are defined. 
Indeed ${\mathbf S}_2$ can be related to  
${\mathbf S}_1$ if we introduce an anti-linear  operator 
${\mathcal C}:{\mathbf S}\rightarrow {\mathbf S}$ defined by
its action on the basis $o^A$, $\iota^A$, $\tilde{o}^A$, $\tilde{\iota}^A$
\begin{equation}
{\mathcal C}(o^A)=-\mbox{i}\tilde{o}^A\;,\quad 
{\mathcal C}(\iota^A)=-\mbox{i}\tilde{\iota}^A\;,\quad 
{\mathcal C}(\tilde o^A)=\mbox{i}o^A\;,\quad
{\mathcal C}(\tilde\iota^A)=\mbox{i}\iota^A.  
\label{eq:C-definition}
\end{equation}
The operator ${\mathcal C}$ has the additional property 
${\mathcal C}^2=-I_{\mathbf S}$ ($I_{\mathbf S}$ is the identity on $\mathbf S$) and therefore it can be used to endow $\mathbf{S}$  with a {\em quaternionic structure}.
Also, we can extend the operator $\mathcal C$ to tensors if in addition to (\ref{eq:C-definition}) we demand that
\begin{equation}
{\mathcal C}(\gamma^a_{\phantom{a}AB})=-\gamma^a_{\phantom{a}AB}.
\label{eq:C-on-gamma}
\end{equation}
When $\mathcal C$ acts on tensors then it becomes an involutive operator and thus 
it is a {\em complex conjugation}.

We can now use the spin tetrad just introduced to construct a basis in ${\mathbf L}$. The way in which this is done is by considering the tensor equivalents of the spinors $o^A\tilde{o}^B$, $\iota^A\tilde{\iota}^B$, $o^A\tilde{\iota}^B$, $\tilde{o}^A\iota^B$, $o^A\iota^B$ and $\tilde{o}^A\tilde{\iota}^B$. These tensor equivalents are
\begin{eqnarray}
&&l{}^{a}\equiv\gamma {}^{a}{}_{A}{}_{B}o{}^{A} \tilde{o}{}^{B}\;,\quad
n{}^{a}\equiv\gamma {}^{a}{}_{A}{}_{B} \iota {}^{A} 
\tilde{\iota}{}^{B}\;,\nonumber\\
&& m{}^{a}\equiv -o{}^{A} \gamma {}^{a}{}_{A}{}_{B}\tilde{\iota}{}^{B}\;, 
\bar{m}{}^{a}\equiv \tilde{o}{}^{A} \gamma {}^{a}{}_{A}{}_{B}\iota{}^{B},\nonumber\\ 
&&u{}^{a}\equiv 2o{}^{B}\gamma{}^{a}{}_{A}{}_{B}\iota{}^{A}= -2\tilde{o}{}^{B}\gamma{}^{a}{}_{A}{}_{B}\tilde{\iota}{}^{A}.
\label{eq:null-pentad}
\end{eqnarray}
From (\ref{eq:C-definition})-(\ref{eq:C-on-gamma}) we deduce that $l^a$, $n^a$ are real with respect to $\mathcal C$ whereas ${\mathcal C}(m^a)=\bar m^a$.
Also using (\ref{eq:null-pentad}) and (\ref{eq:gammasquare}) it is easy to compute the nonvanishing scalar products of the elements of this basis
$$
l^an_a=1\;,\quad m^a\bar{m}_a=-1\;,\quad u^a u_a=-2.
$$
Hence, we deduce that $\{l^a, n^a, m^a,\bar{m}^a,u^a\}$ forms a semi-null {\em pentad}. This is the 5 dimensional analog of the null tetrad used in 4-dimensional Lorentzian geometry and it has a similar relevance. This will be illustrated in the forthcoming sections, where we will perform a number of computations in this basis. 

We adopt a number of general conventions when working with a spin tetrad and its associated semi-null pentad. Suppose that the bases $\mathcal B$ and $\mathcal N$ defined next are respectively a spin tetrad and the semi-null pentad constructed from it
\begin{equation}
{\mathcal B}\equiv\{e^A_{\bf 0},e^A_{\bf 1},e^A_{\bf 2},e^A_{\bf 3}\}\;,\quad
{\mathcal N}\equiv\{e^a_{\bf 1},e^a_{\bf 2},e^a_{\bf 3},e^a_{\bf 4},e^a_{\bf 5}\}.
\label{eq:BN}
\end{equation}
Then we set up the assignments
\begin{eqnarray}
&&e^A_{\bf 0}\equiv o^A\;,\quad e^A_{\bf 1}\equiv \iota^A\;,\quad
e^A_{\bf 2}\equiv \tilde{o}^A\;,\quad 
e^A_{\bf 3}\equiv \tilde{\iota}^A\;,\quad\label{eq:spin-tetrad}\nonumber\\
&&e^a_{\bf 1}\equiv l^a\;,\quad 
e^a_{\bf 2}\equiv n^a\;,\quad  
e^a_{\bf 3}\equiv m^a\;,\quad
e^a_{\bf 4}\equiv \bar m^a\;,\quad e^a_{\bf 5}\equiv u^a. \label{eq:semi-null-pentad}
\end{eqnarray}

The relations written in (\ref{eq:null-pentad}) enable us to obtain right
away the components of $\epsilon_{AB}$ and 
$\gamma^a_{\phantom{a}AB}$ in the bases introduced above. The result is 
\begin{eqnarray}
&&\epsilon_{\bf 01}=1\;,\epsilon_{\bf 23}=-1\;,\quad
\gamma^{\bf 1}_{\phantom{a}\bf 02}=1\;,\gamma^{\bf 2}_{\phantom{a}\bf 13}=1\;,
\gamma^{\bf 3}_{\phantom{a}\bf 03}=-1\;,\quad\gamma^{\bf 4}_{\phantom{a}\bf 12}=-1\;,
\quad\gamma^{\bf 5}_{\phantom{a}\bf 10}=\frac{1}{2}\;,\nonumber\\
&&\gamma^{\bf 5}_{\phantom{a}\bf 23}=-\frac{1}{2}\;, \label{eq:values-of-gamma}
\end{eqnarray}
all the other independent components of $\epsilon_{AB}$ and $\gamma^a_{\phantom{a}AB}$ being zero. Previous result can be written somewhat more invariantly in the form (\ref{eq:epsilondyad}) and 
\begin{eqnarray}
&&\gamma^a_{\phantom{a}AB}=
-2n^ao_{[A}\tilde{o}_{B]}-2l^a\iota_{[A}\tilde{\iota}_{B]}-
2\bar{m}^ao_{[A}\tilde{\iota}_{B]}+2m^a\tilde{o}_{[A}\iota_{B]}
-u^a(\tilde{o}_{[A}\tilde{\iota}_{B]}+o_{[A}\iota_{B]}). 
\label{eq:gamma-expansion}
\end{eqnarray}

\section{Spin structures on a 5-dimensional Lorentzian manifold}
\label{sec:spin-structure}
So far all our considerations have been algebraic in nature, but as is well known one can use these ideas to construct a {\em spin-structure} on a given 5-dimensional Lorentzian manifold. We explain next how this is achieved. Suppose that $(\mathcal{M},g)$ is a 5-dimensional Lorentzian manifold and let $T_p({\mathcal M})$ be the tangent space at a point $p$. This is a vector space which can be endowed with the Lorentzian scalar product $g(\ ,\ )|_p$. Therefore the vector space $T_p({\mathcal M})$ has properties similar to ${\mathbf L}$ and we can introduce a spin space ${\mathbf S}_p$ and a spin structure $\gamma_{aA}^{\phantom{aA}B}|_p$ at each point $p$. 
\begin{definition}[{\bf Spin bundle}]
The union
\begin{equation}
S(\mathcal{M})\equiv\bigcup_{p\in{\mathcal M}} {\mathbf S}_p,
\label{eq:spin-bundle}
\end{equation}
is a vector bundle with the manifold $\mathcal{M}$ as the base space and the group
of linear transformations on $\mathbb C^4$ as the structure group. We will call this vector bundle the \underline{spin bundle} and the sections of $S(\mathcal{M})$ are the contravariant rank-1 spinor fields on ${\mathcal M}$.
\label{def:spin-bundle}
\end{definition}

We can now define the tensor algebras $\mathfrak{T}^r_s(T_p({\mathcal M}))$, 
$\mathfrak{T}^R_S(\mathbf S_p)$ and, by means of a definition similar to (\ref{eq:spin-bundle}) use them to construct vector bundles with $\mathcal M$ as the base manifold. 
These bundles are tensor bundles and we denote each of these tensor bundles by 
$\mathfrak{S}^{r,R}_{s,S}(\mathcal M)$, where the meaning of the labels $r$, $R$, $s$, $S$ is the obvious one.  In general we will suppress these labels and use just the notation 
$\mathfrak{S}(\mathcal M)$ as a generic symbol for these tensor bundles. Sections on $\mathfrak{S}(\mathcal M)$ are written using abstract indices and we follow the same conventions explained for the case of the vector spaces ${\mathbf L}$ and 
${\mathbf S}$. Sections of any of the bundles $\mathfrak{S}^{0,R}_{0,S}(\mathcal M)$
are called {\em spinor fields} or simply {\em spinors}.  
\begin{definition}[{\bf Spin structure on a 5-dimensional manifold}]
If the quantity  $\gamma_{aA}^{\phantom{aA}B}|_p$ varies smoothly on the manifold $\mathcal M$, then one can define a smooth section of the bundle $\mathfrak{S}^{0,1}_{1,1}(\mathcal M)$, denoted by $\gamma_{aA}^{\phantom{aA}B}$. 
When this is the case we call the smooth section $\gamma_{aA}^{\phantom{aA}B}$
a smooth spin structure on the Lorentzian manifold $({\mathcal M},g)$.  
\label{definition:spin-structure}
\end{definition}
A spin structure can be always defined in a neighbourhood of any point 
$p\in{\mathcal M}$, but further topological restrictions on ${\mathcal M}$ are required if the spin structure is to be defined globally. 
A necessary and sufficient condition for the existence
of a spin structure is that the second Stiefel-Whitney class of ${\mathcal M}$ vanishes (see e.g. \cite{NAKAHARA}). 
 
From now on we assume that we are working in a manifold ${\mathcal M}$ admitting a smooth spin structure. Using (\ref{eq:gamma-times-gamma}) one
one can introduce two smooth sections $\epsilon_{AB}$, $\epsilon^{AB}$ 
and use them to raise and lower indices in any spinor field. These sections are defined up to a smooth conformal factor.
The properties shown in eqs. (\ref{eq:epsinverse})-(\ref{eq:kronecker-delta}) hold for $\epsilon_{AB}$, $\epsilon^{AB}$ and the quantity $\delta_A^{\phantom{A}B}$. Also, the algebraic properties shown in theorem \ref{theorem:gamma-properties} and the relations between spinors and tensors explained in subsection \ref{subsection:spinors-tensors} can be carried over to this new context.  

\subsection{Covariant derivatives on $\mathfrak{S}(\mathcal M)$}
\label{subsec:spin-covd}
We turn now to the study of covariant derivatives defined on the tensor bundle $\mathfrak{S}(\mathcal M)$. Let $D_a$ denote such a covariant derivative. Then the operator $D_a$ can act on any quantity with tensor indices and/or spinor indices. As a result, when $D_a$ is restricted to quantities belonging to $\mathfrak{S}^{r,0}_{s,0}(\mathcal M)$ we recover the standard notion of covariant derivative acting on tensor fields of $\mathcal M$. If $D_a$ is restricted to quantities in $\mathfrak{S}^{0,R}_{0,S}(\mathcal M)$ then $D_a$ is the covariant derivative acting on spinor fields. The consequence of this is that the connection coefficients and the curvature of $D_a$ will be divided in two groups: those arising from the tensorial part and those arising from the spinorial part. The group arising from the tensorial part consists of the Christoffel symbols/Ricci rotation coefficients and the Riemann tensor of the covariant derivative restricted to the tangent bundle $T(\mathcal M)$. The group coming from the spinorial part contains the connection components and the curvature tensor of the covariant derivative restricted to the spin bundle $S(\mathcal M)$. We will refer to these as the 
{\em inner connection} and the {\em inner curvature} respectively. See  \cite{ASHTEKAR} for an in-depth discussion of these concepts.

To see how this works in practice, consider a spinor field $\xi^A$. Then the commutation of $D_a$, $D_b$ acting on $\xi^A$ is given by \cite{ASHTEKAR}
\begin{equation}
D{}_{a}
D{}_{b}\xi {}^{A} -D{}_{b}
D{}_{a}\xi {}^{A}=F{}_{b}{}_{a}{}_{B}{}^{A} \xi {}^{B},
\label{eq:ricci-identity-F}
\end{equation}
where we assume that $D_a$ has no torsion (this condition is adopted henceforth for any covariant derivative).
The mixed quantity $F_{abA}^{\phantom{abA}B}$ is the inner curvature mentioned above. It is antisymmetric in the tensorial indices and it fulfils the {\em Bianchi identity} \cite{ASHTEKAR}
\begin{equation}
D_{a}F_{bcA}^{\phantom{bcA}B}+D_{b}F_{caA}^{\phantom{bcA}B}+D_{c}F_{abA}^{\phantom{bcA}B}=0.
\label{eq:F-bianchi-identity} 
\end{equation}

Let now $V\equiv\{e^{A}_{\bf 1},e^{A}_{\bf 2},e^{A}_{\bf 3},e^{A}_{\bf 4}\}$ be a frame on $S(\mathcal M)$ and consider the action of $D_a$ on any element of this frame. The result is
\begin{equation}
D_a e^{A}_{\mathbf B}={\mathcal A}^{A}_{\phantom{A}a{\mathbf B}}. 
\label{eq:D-frame}
\end{equation}
Here and in the following we will use boldface letters to denote basis indices, i.e. indices varying within a range of numbers.
The quantities ${\mathcal A}^{A}_{\phantom{A}a{\mathbf B}}$ are the components of the connection defined by $D_a$ when it is restricted to the vector bundle $S(\mathcal M)$. Traditionally they are regarded as {\em non-tensorial} objects but if we see them as dependent from the frame $\{e^{A}_{\bf 1},e^{A}_{\bf 2},e^{A}_{\bf 3},
e^{A}_{\bf 4}\}$ they can be considered as true tensors \cite{ASHTEKAR}. This is the viewpoint which will be adopted in this work and therefore we shall write
$$
{\mathcal A(D,V)}^A_{\phantom{A}aB},
$$
for the tensor whose components in the frame $V$ yield the quantities appearing in (\ref{eq:D-frame}).
We will call this tensor the {\em inner connection tensor} of $D_a$, and we use a notation which stresses its dependence on the frame $V$. It is possible to obtain a formula for 
$F_{bcA}^{\phantom{bcA}B}$ in terms of the inner connection tensor. The result is 
\cite{ASHTEKAR}
\begin{eqnarray}
&&F{}_{a}{}_{b}{}_{A}{}^{B}=\mathcal{A}(D,V){}^{B}{}_{b}{}_{C}
\mathcal{A}(D,V){}^{C}{}_{a}{}_{A} -\mathcal{A}(D,V){}^{B}{}_{a}{}_{C} \mathcal{A}(D,V){}^{C}{}_{b}{}_{A}-\nonumber\\
&&-\partial{}_{a}\mathcal{A}(D,V){}^{B}{}_{b}{}_{A} + \partial{}_{b}\mathcal{A}(D,V){}^{B}{}_{a}{}_{A},
\label{eq:friemann-formula}
\end{eqnarray}
where $\partial_a$ is any covariant derivative on $\mathfrak{S}(\mathcal M)$ without torsion and curvature.
Now suppose that $v^a$ is a vector field on $T(\mathcal M)$. Then the commutation of $D_a$, $D_b$ on $v^a$ (Ricci identity) yields
$$
D{}_{a}D{}_{b}v{}^{c}-D{}_{b}D{}_{a}v{}^{c}=R{}_{b}{}_{a}{}_{d}{}^{c} v{}^{d}.
$$
The tensor $R{}_{b}{}_{a}{}_{d}{}^{c}$ is the standard Riemann tensor and it fulfils the fa\-mi\-liar first and second Bianchi identities. If we introduce a frame 
$\hat{V}\equiv\{e^a_{\bf 1},e^a_{\bf 2},e^a_{\bf 3},e^a_{\bf 4},e^a_{\bf 5}\}$ we can compute the connection components on it 
$$
D_c e^a_{\mathbf b}=\Gamma^a_{\phantom{a}c{\mathbf b}}.
$$
Again we follow the viewpoint explained above and regard the connection components as the components of a tensor ``attached'' to the frame $\hat{V}$. This tensor is
$$
\Gamma(D,\hat{V})^a_{\phantom{a}bc}.
$$
We shall refer to this tensor as the {\em Christoffel tensor} of $D_a$. Again note the dependency of this tensor on the frame $\hat{V}$. The components of the Christoffel tensor in the frame $\hat{V}$ are known traditionally as the {\em Ricci rotation coefficients} of $D_a$ in that frame. 
If $\hat{V}$ is non-coordinated then $\Gamma(D,\hat{V})^a_{\phantom{a}cb}$ is
not symmetric on its two last indices. If the components $g_{\mathbf ab}$
of the metric in the frame $\hat{V}$ are constants and $D_ag_{bc}=0$
then we have instead the symmetry $\Gamma(D,\hat{V})_{abc}=-\Gamma(D,\hat{V})_{cba}$, where the first index of the Christoffel tensor has been lowered with the metric $g_{ab}$.

\subsection{The spin covariant derivative}
We wish to introduce a particular type of covariant derivative on 
$\mathfrak{S}(\mathcal M)$.
\begin{definition}[\bf Spin covariant derivative]
Suppose that $\mathfrak{S}(\mathcal M)$ admits a spin structure $\gamma_{aA}^{\phantom{aA}B}$. We say that a covariant derivative $D_a$ defined on $\mathfrak{S}(\mathcal M)$ is compatible with the spin structure $\gamma_{aA}^{\phantom{aA}B}$ if it fulfils the
property 
\begin{equation}
D_a\gamma_{bC}^{\phantom{bC}D}=0.
\label{eq:spin-structure-condition}
\end{equation}
The covariant derivative $D_a$ is then called a spin covariant derivative with respect to the spin structure $\gamma_{aA}^{\phantom{aA}B}$.
\label{def:spin-covd}
\end{definition}

\noindent
Acting with such $D_a$ on (\ref{eq:clifford}) gives
\begin{equation}
D_a g_{bc} = 0,
\label{eq:SpinLeviCivita}
\end{equation}
which shows that the restriction of $D_a$ to quantities with tensorial
indices is just the Levi-Civita covariant derivative of $g_{ab}$.
However, condition (\ref{eq:spin-structure-condition}) does not fix
univocally $D_a$ on spinors, and therefore there are many covariant
derivatives which are compatible with a given spin structure. The
freedom originates in the fact that $\epsilon_{AB}$ is defined by
(\ref{eq:gamma-times-gamma}) only up to conformal rescalings.
Differentiating (\ref{eq:gamma-times-gamma}) gives
$$
D_a(\epsilon_{CD}\epsilon^{AB}) = 0,
$$
or equivalently,
$$
D_a \epsilon_{AB}
= \frac{1}{4}\left(\epsilon^{CD}D_{a}\epsilon_{CD}\right)\epsilon_{AB}
= \epsilon_{AB}D_{a}Y\;,\quad Y\equiv\frac{1}{4}\log\det\epsilon.
$$
Hence, $D_a$ is of Weyl type with respect to the metric $\epsilon_{AB}$,
but it is always possible to switch to another compatible spin derivative
which is of Levi-Civita type:

\begin{theorem}
There is one and only one spin covariant derivative 
$\nabla_a$ on $\mathfrak{S}(\mathcal M)$ with respect to the spin structure $\gamma_{aA}^{\phantom{aA}B}$ which fulfils the property
\begin{equation}
\nabla_a\epsilon_{AB}=0.
\label{eq:cd-epsilon}
\end{equation}
\label{theorem:spincd-unique}
\end{theorem}

\noindent
\Proof
Condition (\ref{eq:SpinLeviCivita}) determines the action of $\nabla_a$
on tensors and hence the components of the Christoffel tensor of $\nabla_a$
in any frame are the familiar connection components of the Levi-Civita
covariant derivative.
We now need to show that there is a frame in which the inner connection tensor gets also fixed.
Let us work in a spin tetrad whose properties are those described in (\ref{eq:epsilondyad})-(\ref{eq:dyadnormalization}) and construct from it a null pentad in the way shown in (\ref{eq:null-pentad}). 
We also need to introduce the frame derivations of $\mathcal N$ which are
\begin{equation}
D\equiv l^a\nabla_a\;,\quad
\Delta\equiv n^a\nabla_a\;,\quad
\delta\equiv m^a\nabla_a\;,\quad
\bar\delta\equiv\bar m^a\nabla_a\;,\quad\mathcal{D}\equiv u^a\nabla_a.
\label{eq:frame-derivations}
\end{equation}
The operators $D$, $\Delta$, $\delta$ and $\bar\delta$ correspond to the standard 
Newman-Penrose frame derivations used in 4-dimensional Lorentz geometry whereas ${\mathcal D}$ has to be added in order to work in five dimensions.

Now we take the conditions $\nabla_a\epsilon_{AB}=0$, $\nabla_a\gamma^b_{\phantom{b}AB}=0$ and expand them in the spin tetrad $\mathcal{B}$, 
and the semi-null pentad $\mathcal{N}$ (see (\ref{eq:spin-tetrad})-(\ref{eq:semi-null-pentad})). 
The derivatives of the components of $\epsilon_{AB}$ and $\gamma^b_{\phantom{b}AB}$ are
\begin{eqnarray}
&&\nabla_{\mathbf a}\epsilon_{{\mathbf A}{\mathbf B}}=
{\mathcal A}(\nabla,\mathcal B)^{\mathbf C}_{\phantom{\mathbf C}{\mathbf a}{\mathbf B}}\epsilon_{{\mathbf A}{\mathbf C}}+{\mathcal A}(\nabla,\mathcal B)^{\mathbf C}_{\phantom{\mathbf C}{\mathbf a}{\mathbf A}}\epsilon_{{\mathbf C}{\mathbf 
B}}\label{eq:expand-nabla-epsilon},\\
&&\nabla_{\mathbf a}\gamma^{\mathbf b}_{\phantom{\mathbf b}{\mathbf A}{\mathbf B}}=
{\mathcal A}(\nabla,\mathcal B)^{\mathbf C}_{\phantom{\mathbf C}{\mathbf a}{\mathbf B}}\gamma^{\mathbf b}_{\phantom{\mathbf b}{\mathbf A}{\mathbf C}}+{\mathcal A}(\nabla,\mathcal B)^{\mathbf C}_{\phantom{\mathbf C}{\mathbf a}{\mathbf A}}\gamma^{\mathbf b}_{\phantom{\mathbf b}{\mathbf C}
{\mathbf B}}-\Gamma(\nabla,\mathcal N)^{\mathbf b}_{\phantom{\mathbf b}{\mathbf a}{\mathbf c}}\gamma^{\mathbf c}_{\phantom{\mathbf c}{\mathbf A}{\mathbf B}},
\label{eq:expand-nabla-gamma}
\end{eqnarray}
where $\nabla_{\bf 1}=D$, $\nabla_{\bf 2}=\Delta$, $\nabla_{\bf 3}=\delta$, $\nabla_{\bf 4}=\bar\delta$ and $\nabla_{\bf 5}=\mathcal{D}$ are just  
the frame differentiations defined in (\ref{eq:frame-derivations}). Since $\epsilon_{{\mathbf A}{\mathbf B}}$, $\gamma^{\mathbf b}_{\phantom{\mathbf b}{\mathbf A}{\mathbf B}}$ are constants for any value of the basis indices, we deduce that the left hand side of (\ref{eq:expand-nabla-epsilon})-(\ref{eq:expand-nabla-gamma}) is zero. The values of $\gamma^{\mathbf b}_{\phantom{\mathbf b}{\mathbf A}{\mathbf B}}$ and $\epsilon_{{\mathbf A}{\mathbf B}}$ are known
(see (\ref{eq:gamma-expansion}) and (\ref{eq:epsilondyad})) and $\Gamma(\nabla,\mathcal N)^{\mathbf b}_{\phantom{\mathbf b}{\mathbf a}{\mathbf c}}$ are the Ricci rotation coefficients of the Levi-Civita connection of $g_{ab}$ in the semi-null pentad ${\mathcal N}$ so they are also fixed (they are a set of 50 independent scalar quantities, because of the symmetry
$\Gamma(\nabla,\mathcal{N})_\mathbf{abc}=
-\Gamma(\nabla,\mathcal{N})_\mathbf{cba}$, which leaves 5 times 10
antisymmetric pairs). Condition (\ref{eq:expand-nabla-epsilon})
contains 30 independent equations (5 times 6 antisymmetric pairs) and hence it reduces the number of independent scalars 
$A(\nabla,\mathcal B)^{\mathbf C}_{\phantom{\mathbf C}{\mathbf a}{\mathbf B}}$ down to 80-30=50. In other words, lowering the first index of the inner connection tensor we see that it is symmetric: $\mathcal{A}(\nabla,\mathcal{B})_\mathbf{AcB}=\mathcal{A}(\nabla,\mathcal{B})_\mathbf{BcA}$. Hence (\ref{eq:expand-nabla-gamma}) can be regarded as a linear system in the 50 scalars of the set of components  
$A(\nabla,\mathcal B)^{\mathbf C}_{\phantom{\mathbf C}{\mathbf a}{\mathbf B}}$ taken as independent. The linear system can be solved explicitly by writing out (\ref{eq:expand-nabla-gamma}) and one finds that it is possible to obtain
a unique value for these independent quantities in terms of the 50 independent Ricci rotation coefficients. Thus, having determined all the scalars $\Gamma(\nabla,\mathcal N)^{\mathbf b}_{\phantom{\mathbf b}{\mathbf a}{\mathbf c}}$ and $A(\nabla,\mathcal B)^{\mathbf C}_{\phantom{\mathbf C}{\mathbf a}{\mathbf B}}$  we conclude that $\nabla_a$ itself is completely determined. \qed 

From now on the symbol $\nabla_a$ will be reserved for the covariant derivative introduced in the previous theorem. Therefore when we speak of the spin covariant derivative, we will mean the spin covariant derivative $\nabla_a$ which is compatible with $\epsilon_{AB}$. Hence the Riemann tensor 
$R_{abcd}$ will be always the Riemann tensor of this spin covariant derivative (which as explained above is just the Riemann tensor of the Levi-Civita connection of $g_{ab}$). 
To shorten certain expressions, we introduce the differential operator
\begin{equation}
\nabla_{AB}\equiv \gamma^a_{\phantom{a}AB}\nabla_a.
\label{eq:define-nabla-2index} 
\end{equation}
From this definition we obtain the following straightforward properties
\begin{equation}
\nabla_{[AB]}=\nabla_{AB}\;,\quad \nabla^A_{\phantom{A}A}=0. 
\label{eq:nabla-2index-properties}
\end{equation}

The linear relation which gives the 50 independent inner connection components in terms of the 50 independent Ricci rotation coefficients can be inverted yielding
the independent values of  $\Gamma(\nabla,\mathcal N)^{\mathbf b}_{\phantom{\mathbf b}{\mathbf a}{\mathbf c}}$ in terms of the independent values of $A(\nabla,\mathcal B)^{\mathbf C}_{\phantom{\mathbf C}{\mathbf a}{\mathbf B}}$.  We write this symbolically in the form
\begin{equation}
\Gamma^{ind}(\nabla,\mathcal N)^{\mathbf b}_{\phantom{\mathbf b}{\mathbf a}{\mathbf c}}\rightarrow A^{ind}(\nabla,\mathcal B)^{\mathbf C}_{\phantom{\mathbf C}{\mathbf a}{\mathbf B}},
\label{eq:gammaind-to-aind}
\end{equation}
where $\Gamma^{ind}(\nabla,\mathcal N)^{\mathbf b}_{\phantom{\mathbf b}{\mathbf a}{\mathbf c}}$, $A^{ind}(\nabla,\mathcal B)^{\mathbf C}_{\phantom{\mathbf C}{\mathbf a}{\mathbf B}}$ denote, respectively, the independent Ricci rotation coefficients and inner connection components. It is possible to reduce further the number of independent Ricci rotation coefficients if we take into account that some of them are complex numbers. For example the components 
$\Gamma(\nabla,\mathcal N)^\mathbf{1}_{\phantom{1}\mathbf 13}$ and 
$\Gamma(\nabla,\mathcal N)^\mathbf{1}_{\phantom{1}\mathbf 14}$ are both in the set $\Gamma^{ind}(\nabla,\mathcal N)^{\mathbf b}_{\phantom{\mathbf b}{\mathbf a}{\mathbf c}}$ but they fulfil the relation
\begin{equation}
\overline{\Gamma(\nabla,\mathcal N)^\mathbf{1}_{\phantom{1}\mathbf 13}}=\Gamma(\nabla,\mathcal N)^\mathbf{1}_{\phantom{1}\mathbf 14},
\label{eq:dagger-example}
\end{equation}
which comes from the fact that $m^a$ and $\bar m^a$ are complex conjugate of each other. Therefore if we compute all the independent relations of this type and use (\ref{eq:gammaind-to-aind}) on them we will obtain relations among the inner connection components similar to (\ref{eq:dagger-example}). We explain in next subsection how to take advantage of this fact in practical computations.

\subsubsection{The spin coefficients}
For practical computations one takes the independent components of the inner connection which have been obtained and then introduces specific symbols to denote each scalar component. We show next which are these independent components and their assigned names. We start with
\begin{eqnarray}
&&{\mathcal A}(\nabla,\mathcal B){}^{{\bf 1}}{}_{{\bf 4}}{}_{{\bf 1}}=-\alpha\;,\quad 
{\mathcal A}(\nabla,\mathcal B){}^{{\bf 1}}{}_{{\bf 3}}{}_{{\bf 1}}=-\beta \;,\quad 
{\mathcal A}(\nabla,\mathcal B){}^{{\bf 1}}{}_{{\bf 2}}{}_{{\bf 1}}=-\gamma \;,\quad 
{\mathcal A}(\nabla,\mathcal B){}^{{\bf 1}}{}_{{\bf 1}}{}_{{\bf 1}}=-\epsilon \;,\quad\nonumber\\ 
&&{\mathcal A}(\nabla,\mathcal B){}^{{\bf 1}}{}_{{\bf 1}}{}_{{\bf 0}}=-\kappa \;,\quad 
{\mathcal A}(\nabla,\mathcal B){}^{{\bf 0}}{}_{{\bf 4}}{}_{{\bf 1}}=\lambda\;,\quad 
{\mathcal A}(\nabla,\mathcal B){}^{{\bf 0}}{}_{{\bf 3}}{}_{{\bf 1}}=\mu\;,\quad 
 {\mathcal A}(\nabla,\mathcal B){}^{{\bf 0}}{}_{{\bf 2}}{}_{{\bf 1}}=\nu\;,\quad\nonumber\\
&& {\mathcal A}(\nabla,\mathcal B){}^{{\bf 0}}{}_{{\bf 1}}{}_{{\bf 1}}=\pi \
\;,\quad {\mathcal A}(\nabla,\mathcal B){}^{{\bf 1}}{}_{{\bf 4}}{}_{{\bf 0}}=-\rho\;,\quad 
{\mathcal A}(\nabla,\mathcal B){}^{{\bf 1}}{}_{{\bf 3}}{}_{{\bf 0}}=-\sigma\;,\quad 
{\mathcal A}(\nabla,\mathcal B){}^{{\bf 1}}{}_{{\bf 2}}{}_{{\bf 0}}=-\tau.\nonumber\\
\label{eq:define-np}
\end{eqnarray}
These scalars are in fact the twelve complex Newman-Penrose spin coefficients which appear in the spinor calculus of 4-dimensional Lorentzian geometry. Since the 4-dimensional Lorentzian geometry can be seen as a restriction of the  5-dimensional one it is then reasonable that the Newman-Penrose spin coefficients also appear in our context. However, in order to work with generic 5-dimensional spacetimes we need to add more spin coefficients to the set (\ref{eq:define-np}) which is what is done next. The ``new'' spin coefficients are 
\begin{eqnarray}
&&{\mathcal A}(\nabla,\mathcal B){}^{{\bf 0}}{}_{{\bf 5}}{}_{{\bf 1}}=\zeta \;, 
{\mathcal A}(\nabla,\mathcal B){}^{{\bf 1}}{}_{{\bf 5}}{}_{{\bf 0}}=\eta \;, 
{\mathcal A}(\nabla,\mathcal B){}^{{\bf 1}}{}_{{\bf 5}}{}_{{\bf 1}}=\theta \;, 
{\mathcal A}(\nabla,\mathcal B){}^{{\bf 2}}{}_{{\bf 1}}{}_{{\bf 0}}=\chi \;,\nonumber \\
&&{\mathcal A}(\nabla,\mathcal B){}^{{\bf 2}}{}_{{\bf 2}}{}_{{\bf 0}}=\omega\;, 
{\mathcal A}(\nabla,\mathcal B){}^{{\bf 2}}{}_{{\bf 3}}{}_{{\bf 0}}=\phi \;, 
{\mathcal A}(\nabla,\mathcal B){}^{{\bf 2}}{}_{{\bf 3}}{}_{{\bf 1}}=\xi\;, 
{\mathcal A}(\nabla,\mathcal B){}^{{\bf 2}}{}_{{\bf 4}}{}_{{\bf 0}}=\upsilon\;,\nonumber\\ 
&&{\mathcal A}(\nabla,\mathcal B){}^{{\bf 2}}{}_{{\bf 5}}{}_{{\bf 0}}=\psi \;, 
{\mathcal A}(\nabla,\mathcal B){}^{{\bf 3}}{}_{{\bf 3}}{}_{{\bf 0}}=\varsigma.
\label{eq:define-complexcoefficients}
\end{eqnarray}
This is a set of ten complex quantities. In addition we need to include a set of six real spin coefficients
\begin{eqnarray}
&&{\mathcal A}(\nabla,\mathcal B){}^{{\bf 2}}{}_{{\bf 1}}{}_{{\bf 1}}=\mathfrak{a}\;,\quad 
{\mathcal A}(\nabla,\mathcal B){}^{{\bf 2}}{}_{{\bf 2}}{}_{{\bf 1}}=\mathfrak{b}\;,\quad 
{\mathcal A}(\nabla,\mathcal B){}^{{\bf 2}}{}_{{\bf 5}}{}_{{\bf 1}}=\mathfrak{c}\;,\quad\nonumber\\ 
&&{\mathcal A}(\nabla,\mathcal B){}^{{\bf 3}}{}_{{\bf 1}}{}_{{\bf 0}}=\mathfrak{d}\;,\quad 
{\mathcal A}(\nabla,\mathcal B){}^{{\bf 3}}{}_{{\bf 2}}{}_{{\bf 0}}=\mathfrak{e}\;,\quad 
{\mathcal A}(\nabla,\mathcal B){}^{\bf 3}{}_{{\bf 5}}{}_{{\bf 0}}=\mathfrak{f}.
\label{eq:define-realcoefficients}
\end{eqnarray}
Therefore we have the twelve complex Newman-Penrose spin coefficients, the ten complex spin coefficients of (\ref{eq:define-complexcoefficients}) and the six real spin coefficients of (\ref{eq:define-realcoefficients}). They add up to 50 independent real quantities as they should. We will call the spin coefficients defined in (\ref{eq:define-complexcoefficients})-(\ref{eq:define-realcoefficients}) the 5-dimensional spin coefficients.

Using the information of (\ref{eq:define-np})-(\ref{eq:define-realcoefficients}) we can compute the action of the operators defined in (\ref{eq:frame-derivations}) on the spin tetrad elements. The result is 
\begin{eqnarray}
&&Do{}_{A}=\epsilon o{}_{A}  + \mathfrak {d} 
\tilde{\iota }{}_{A} -\kappa\iota {}_{A}   + 
\chi\tilde{o}{}_{A} \;,\quad \Delta o{}_{A}=\gamma o{}_{A} 
  + \mathfrak {e} \tilde{\iota }{}_{A}-\tau\iota{}_{A} 
  + \omega\tilde{o}{}_{A}  \;,\quad\nonumber\\ 
&&\delta o{}_{A}=\beta o{}_{A}-\sigma\iota {}_{A}  + 
\varsigma\tilde{\iota }{}_{A}   + \phi\tilde{o}{}_{A} \;,\quad 
\bar\delta o{}_{A}=\alpha o{}_{A} -\rho\iota {}_{A}  
+ \bar\varsigma\tilde{\iota }{}_{A} +\upsilon\tilde{o}{}_{A}  \;,\quad\nonumber\\ &&{\mathcal D} o{}_{A}=-\theta o{}_{A}   + \eta  \iota {}_{A} + \mathfrak{f} \
\tilde{\iota}{}_{A} + \psi\tilde{o}{}_{A}  \;,\quad 
D\iota {}_{A}=\mathfrak {a} \tilde{o}{}_{A} -\epsilon\iota {}_{A} + 
\pi o{}_{A}  -\bar\chi\tilde{\iota}{}_{A}\;,\quad\nonumber\\ 
&&\Delta\iota {}_{A}=\mathfrak {b}\tilde{o}{}_{A} -\gamma  
\iota {}_{A} + \nu o{}_{A}  -\bar\omega\tilde{\iota }{}_{A}\;,\quad 
\delta \iota{}_{A}=-\beta  \iota {}_{A} +\mu o{}_{A} + 
\xi\tilde{o}{}_{A}  -\bar\upsilon\tilde{\iota }{}_{A} \;,\quad\nonumber\\ 
&&\bar\delta \iota {}_{A}=-\alpha \iota {}_{A} + \lambda o{}_{A}  + \bar\xi\tilde{o}{}_{A}-\bar\phi\tilde{\iota }{}_{A} \;,\quad 
{\mathcal D} \iota{}_{A}=\mathfrak {c} \tilde{o}{}_{A} + \zeta o{}_{A}  + 
\theta \iota{}_{A} -\bar\psi\tilde{\iota }{}_{A}.
\label{eq:differential-spin-tetrad}
\end{eqnarray}
From this set we can obtain a similar set of equations for $\tilde{o}$,
$\tilde{\iota}$  if we use the operator $\mathcal C$. Again (\ref{eq:differential-spin-tetrad}) generalises the expression which gives the action of the Newman-Penrose frame differentiations on the elements of a spin dyad when working in 4-dimensional Lorentzian geometry (see eq. (4.5.26) of \cite{PENROSERINDLER1}).

Using the spin coefficients introduced above we can write out (\ref{eq:gammaind-to-aind}) explicitly. The result is
\begin{eqnarray}
&&\Gamma(\nabla,\mathcal N){}^{\bf 2}{}_{\bf 1}{}_{\bf 2}=-\epsilon-\bar\epsilon\;,\quad
\Gamma(\nabla,\mathcal N){}^{\bf 2}{}_{\bf 2}{}_{\bf 2}=-\gamma -\bar\gamma\;,\quad
\Gamma(\nabla,\mathcal N){}^{\bf 2}{}_{\bf 3}{}_{\bf 2}=-\bar\alpha-\beta,\nonumber\\ 
&&\Gamma(\nabla,\mathcal N){}^{\bf 2}{}_{\bf 5}{}_{\bf 2}=\theta  + \bar\theta\;,\quad\Gamma(\nabla,\mathcal N){}^{\bf 3}{}_{\bf 1}{}_{\bf 1}=\bar\kappa\;,\quad 
\Gamma(\nabla,\mathcal N){}^{\bf 3}{}_{\bf 1}{}_{\bf 2}=-\pi\;,\quad  
\Gamma(\nabla,\mathcal N){}^{\bf 3}{}_{\bf 2}{}_{\bf 1}=\bar\tau,\nonumber\\ 
&&\Gamma(\nabla,\mathcal N){}^{\bf 3}{}_{\bf 2}{}_{\bf 2}=-\nu\;,\quad  \Gamma(\nabla,\mathcal N){}^{\bf 3}{}_{\bf 3}{}_{\bf 1}=\bar\rho\;,\quad  
\Gamma(\nabla,\mathcal N){}^{\bf 3}{}_{\bf 3}{}_{\bf 2}=-\mu\;,\quad  
\Gamma(\nabla,\mathcal N){}^{\bf 3}{}_{\bf 4}{}_{\bf 1}=\bar\sigma,\nonumber\\  
&&\Gamma(\nabla,\mathcal N){}^{\bf 3}{}_{\bf 4}{}_{\bf 2}=-\lambda\;,\quad  
\Gamma(\nabla,\mathcal N){}^{\bf 3}{}_{\bf 5}{}_{\bf 1}=-\bar\eta\;,\quad \Gamma(\nabla,\mathcal N){}^{\bf 3}{}_{\bf 5}{}_{\bf 2}=-\zeta\;,\quad  
\Gamma(\nabla,\mathcal N){}^{\bf 4}{}_{\bf 1}{}_{\bf 4}=-\epsilon + \bar\epsilon,\nonumber\\  
&&\Gamma(\nabla,\mathcal N){}^{\bf 4}{}_{\bf 2}{}_{\bf 4}=-\gamma  + \bar\gamma\;,\quad
\Gamma(\nabla,\mathcal N){}^{\bf 4}{}_{\bf 3}{}_{\bf 4}=\bar\alpha-\beta,\nonumber\\ 
&&\Gamma(\nabla,\mathcal N){}^{\bf 4}{}_{\bf 5}{}_{\bf 4}=\theta -\bar\theta\;,\quad
\Gamma(\nabla,\mathcal N){}^{\bf 5}{}_{\bf 1}{}_{\bf 1}=\mathfrak{d}\;,\quad  
\Gamma(\nabla,\mathcal N){}^{\bf 5}{}_{\bf 1}{}_{\bf 2}=-\mathfrak{a}\;,\quad
\Gamma(\nabla,\mathcal N){}^{\bf 5}{}_{\bf 1}{}_{\bf 3}=\chi,\nonumber\\ 
&&\Gamma(\nabla,\mathcal N){}^{\bf 5}{}_{\bf 3}{}_{\bf 2}=-\xi\;,\quad  
\Gamma(\nabla,\mathcal N){}^{\bf 5}{}_{\bf 3}{}_{\bf 3}=\phi\;,\quad  
\Gamma(\nabla,\mathcal N){}^{\bf 5}{}_{\bf 3}{}_{\bf 4}=\bar\upsilon\;,\quad  
\Gamma(\nabla,\mathcal N){}^{\bf 5}{}_{\bf 5}{}_{\bf 1}=\mathfrak{f},\nonumber\\ 
&&\Gamma(\nabla,\mathcal N){}^{\bf 5}{}_{\bf 2}{}_{\bf 1}=\mathfrak{e}\;,\quad  
\Gamma(\nabla,\mathcal N){}^{\bf 5}{}_{\bf 2}{}_{\bf 2}=-\mathfrak{b}\;,\quad  
\Gamma(\nabla,\mathcal N){}^{\bf 5}{}_{\bf 2}{}_{\bf 3}=\omega\;,\quad  
\Gamma(\nabla,\mathcal N){}^{\bf 5}{}_{\bf 3}{}_{\bf 1}=\varsigma,
\nonumber\\
&&\Gamma(\nabla,\mathcal N){}^{\bf 5}{}_{\bf 5}{}_{\bf 2}=-\mathfrak{c}\;,\quad 
\Gamma(\nabla,\mathcal N){}^{\bf 5}{}_{\bf 5}{}_{\bf 3}=\psi.
\label{eq:riccirotation-to-spin-coeffs} 
\end{eqnarray}
We note that one needs to compute the complex conjugate of some of the above equations in order to obtain the value of all the 50 independent Ricci rotation coefficients. 

\section{The curvature spinors}
\label{section:curvature-spinors}
In this section we compute the spinor counterpart of the Riemann tensor $R_{abcd}$ and we decompose the resulting spinor into {\em irreducible parts}. The result of such a decomposition yields the {\em curvature spinors} which completely characterise the Riemann tensor. The situation is completely analogous to the case of spinor calculus in 4-dimensional Lorentzian geometry and we will obtain different curvature spinors for each of the irreducible parts in which the Riemann tensor is decomposed (Weyl, traceless Ricci and scalar curvature). Some similarities with the curvature spinors of the spinor calculus of 4-dimensional Lorentzian Geometry can be expected. For example we will find a totally symmetric {\em Weyl spinor} but important differences with the 4-dimensional case are also present as we will discuss.

The starting point is the Ricci identity written for an arbitrary vector $v^c$ 
$$
\nabla{}_{a}\nabla{}_{b}v{}^{c}-\nabla{}_{b}\nabla{}_{a}v{}^{c}=
R{}_{b}{}_{a}{}_{d}{}^{c}v{}^{d}.
$$ 
We replace in this expression the tensor $v^a$ by (see (\ref{eq:tensor=spinor}))
$$
v^a=\gamma^a_{\phantom{a}AB}v^{AB}.
$$
The Ricci identity becomes
\begin{equation}
\gamma {}^{c}{}_{A}{}_{B}( \nabla{}_{a}\nabla{}_{b}v{}^{A}{}^{B} 
-\nabla{}_{b}\nabla{}_{a}v{}^{A}{}^{B})=
R{}_{b}{}_{a}{}_{d}{}^{c} \gamma{}^{d}{}_{A}{}_{B}v{}^{A}{}^{B},
\label{eq:ricci-identity-v}
\end{equation}
where the condition $\nabla_a\gamma^b_{\phantom{b}AB}=0$ was used. Next we use the Ricci identity (\ref{eq:ricci-identity-F}) particularised for $v^{AB}$ and $\nabla_a$, which is 
$$
\nabla{}_{a}\nabla{}_{b}v{}^{A}{}^{B}-\nabla{}_{b}\nabla{}_{a}
v{}^{A}{}^{B}=F{}_{b}{}_{a}{}_{C}{}^{B}v{}^{A}{}^{C} + 
F{}_{b}{}_{a}{}_{C}{}^{A}v{}^{C}{}^{B}
$$
Using this in (\ref{eq:ricci-identity-v}) we get after some algebra.
$$
v{}_{A}{}_{B} (F{}_{a}{}_{b}{}^{B}{}^{C}\gamma{}^{c}{}_{C}{}^{A} -F{}_{a}{}_{b}{}^{A}{}^{C} \gamma{}^{c}{}_{C}{}^{B}- R{}_{a}{}_{b}{}^{c}{}_{d}\gamma{}^{d}{}^{A}{}^{B})=0.
$$
Here the spinor $v^{AB}$ can be regarded as an arbitrary antisymmetric spinor. Hence
$$
F{}_{a}{}_{b}{}^{B}{}^{C}\gamma{}^{c}{}_{C}{}^{A} -F{}_{a}{}_{b}{}^{A}{}^{C} \gamma{}^{c}{}_{C}{}^{B}- R{}_{a}{}_{b}{}^{c}{}_{d}\gamma{}^{d}{}^{A}{}^{B}=0.
$$
We multiply both sides of this expression by $\gamma_{fAB}$ and use (\ref{eq:gamma2-contracted}) getting
$$
2 R{}_{a}{}_{b}{}^{c}{}_{f}-F{}_{a}{}_{b}{}^{B}{}^{C}\gamma {}^{c}{}^{A}{}_{C}
\gamma {}_{f}{}_{A}{}_{B} + F{}_{a}{}_{b}{}^{A}{}^{C} 
\gamma {}^{c}{}^{B}{}_{C}\gamma {}_{f}{}_{A}{}_{B}=0,
$$
from which we obtain 
$$
R{}_{a}{}_{b}{}_{c}{}_{f}=
F{}_{a}{}_{b}{}^{A}{}^{B}\gamma{}_{c}{}_{B}{}^{C}\gamma {}_{f}{}_{A}{}_{C}.
$$
This last expression can be more conveniently written if we use (\ref{eq:productof2gammas})
with the result
\begin{equation}
R{}_{a}{}_{b}{}_{c}{}_{f}=-F{}_{a}{}_{b}{}^{A}{}^{B}
G{}_{c}{}_{f}{}_{A}{}_{B}.
\label{eq:RiemannToFRiemann}
\end{equation}
We introduce now the spinor $X_{ABCD}$ by means of the relation 
\begin{equation}
X{}_{C}{}_{D}{}_{A}{}_{B}\equiv F{}_{a}{}_{b}{}_{A}{}_{B}
G{}^{a}{}^{b}{}_{C}{}_{D}.
\label{eq:X-definition}
\end{equation}
Clearly $X_{ABCD}$ is symmetric in the last pair of indices. Previous relation can be inverted using (\ref{eq:Gsquares}) yielding
\begin{equation}
F{}_{c}{}_{d}{}_{A}{}_{B}=\frac{1}{2} 
G{}_{c}{}_{d}{}_{C}{}_{D} X{}^{C}{}^{D}{}_{A}{}_{B},
\label{eq:FRiemanToX}
\end{equation}
which replaced back in (\ref{eq:RiemannToFRiemann}) leads to
\begin{equation}
 R{}_{a}{}_{b}{}_{c}{}_{f}=-\frac{1}{2} G{}_{a}{}_{b}{}_{A}{}_{B} 
G{}_{c}{}_{f}{}_{C}{}_{D} X{}^{A}{}^{B}{}^{C}{}^{D}.
\label{eq:RiemannToXSpinor}
\end{equation}
A straightforward consequence of this relation is that we can choose $X_{ABCD}$ invariant under the interchange of the first and the second pair of indices (and hence symmetric also in the first pair of indices). Thus
\begin{equation}
X_{ABCD}=X_{(AB)CD}=X_{AB(CD)}=X_{CDAB}.
\label{eq:symmetries-of-X}
\end{equation}
The spinor $X_{ABCD}$ can be regarded as the spinor counterpart of the Riemann tensor.
We can extract further information out of it by finding its decomposition into {\em irreducible parts} under the action of the general linear group. The computation of this decomposition for spinors in 5-dimensional Lorentz geometry is far more difficult than for spinors in 4-dimensional Lorentz geometry because in the former case we need to use the general techniques to decompose a tensor into irreducible parts under the general linear transformation group. It falls well beyond the scope of this paper to explain how this decomposition is performed in general (an account of this can be found in e.g. \cite{TUNG}) and we will limit ourselves to explaining how the procedure works in the particular case of the spinor $X_{ABCD}$.

\subsection{Irreducible decomposition of the spinor $X_{ABCD}$}
To obtain the irreducible decomposition of $X_{ABCD}$ we shall proceed in two steps. In the first step we find the decomposition of $X_{ABCD}$ in parts invariant under the {\em trace operation}. In the second step we take each term of this decomposition and 
split it into a sum of terms invariant under the action of the {\em symmetric group} (this is a generic procedure to obtain the irreducible decomposition of any tensor).  

There is a general algorithm for the decomposition of any tensor as a linear combination of traceless tensors multiplied by Kronecker deltas \cite{KRUPKA}. Unfortunately, there is no general formula to get the coefficients of such linear combination, which must be obtained by solving linear systems which rapidly grow in size with the rank of the original tensor. The decomposition for a rank-4 spinor $X^{AB}_{\phantom{AB}CD}$ in dimension 4 reads \cite{KRUPKA}
\begin{eqnarray*}
X{}^{A}{}^{B}{}_{C}{}_{D}&=&W{}^{A}{}^{B}{}_{C}{}_{D} +
\frac{1}{24} \delta {}^{A}{}_{D} (2
X{}^{B}{}^{F}{}_{C}{}_{F}-X{}^{B}{}^{F}{}_{F}{}_{C} -7
X{}^{F}{}^{B}{}_{C}{}_{F} + 2 X{}^{F}{}^{B}{}_{F}{}_{C})
\nonumber\\ &&
+ \frac{1}{24} \delta {}^{A}{}_{C} 
(-X{}^{B}{}^{F}{}_{D}{}_{F}+2 X{}^{B}{}^{F}{}_{F}{}_{D}+2 X{}^{F}{}^{B}{}_{D}{}_{F} -7 X{}^{F}{}^{B}{}_{F}{}_{D})
\nonumber\\ &&
+\frac{1}{24}\delta {}^{B}{}_{C}(2 
X{}^{A}{}^{F}{}_{D}{}_{F} -7 X{}^{A}{}^{F}{}_{F}{}_{D}- 
X{}^{F}{}^{A}{}_{D}{}_{F} + 2 X{}^{F}{}^{A}{}_{F}{}_{D})
\nonumber\\ &&
+\frac{1}{24} \delta {}^{B}{}_{D}
(-7X{}^{A}{}^{F}{}_{C}{}_{F} + 2 X{}^{A}{}^{F}{}_{F}{}_{C} +
2 X{}^{F}{}^{A}{}_{C}{}_{F} -X{}^{F}{}^{A}{}_{F}{}_{C})
\nonumber\\ &&
+\frac{1}{30}\delta {}^{B}{}_{D}\delta {}^{A}{}_{C} (-3
X{}^{F}{}^{H}{}_{F}{}_{H} + 2 X{}^{F}{}^{H}{}_{H}{}_{F})
\nonumber\\ &&
+\frac{1}{30} \delta {}^{B}{}_{C}\delta {}^{A}{}_{D} (2X{}^{F}{}^{H}{}_{F}{}_{H} -3X{}^{F}{}^{H}{}_{H}{}_{F}),
\end{eqnarray*}
where $W^{AB}_{\phantom{AB}CD}$ is a completely traceless spinor, namely
$$
W^{CB}_{\phantom{CD}CD}=W^{AC}_{\phantom{CD}CD}=W^{DB}_{\phantom{CD}CD}=
W^{AD}_{\phantom{CD}CD}=0.
$$
If we use now the symmetries (\ref{eq:symmetries-of-X}) the decomposition of $X^{AB}_{\phantom{AB}CD}$ found above becomes (we lower all indices)
\begin{eqnarray}
X{}_{A}{}_{B}{}_{C}{}_{D} &=& W{}_{C}{}_{D}{}_{A}{}_{B}
-\frac{1}{6} (X{}_{B}{}^{F}{}_{D}{}_{F}\epsilon{}_{A}{}_{C}+X{}_{B}{}^{F}{}_{C}{}_{F} \
\epsilon {}_{A}{}_{D}+X{}_{A}{}^{F}{}_{D}{}_{F} \epsilon {}_{B}{}_{C}+
X{}_{A}{}^{F}{}_{C}{}_{F} \epsilon {}_{B}{}_{D})\nonumber\\
&&-\frac{X{}^{F}{}^{H}{}_{F}{}_{H}}{30}(\epsilon{}_{A}{}_{D} \epsilon {}_{B}{}_{C} + 
\epsilon{}_{A}{}_{C} \epsilon {}_{B}{}_{D}).
\label{eq:decompose-X-trace}
\end{eqnarray}
From previous expression is easy to deduce that, besides it being completely traceless, the spinor $W_{ABCD}$ has the same symmetries as $X_{ABCD}$. 

Now we need to compute the decomposition into irreducible parts of each term of (\ref{eq:decompose-X-trace}). We start by noting the formula
\begin{equation}
X{}_{A}{}^{F}{}_{C}{}_{F}=\Sigma{}_{C}{}_{A}+\frac{1}{4}X{}^{B}{}^{D}{}_{B}{}_{D} \epsilon{}_{C}{}_{A},
\label{eq:decompose-X-step2}
\end{equation}
which is the decomposition of the quantity $X{}_{A}{}^{F}{}_{C}{}_{F}$ in parts invariant under the trace. Here the spinor $\Sigma_{CA}$ is traceless, $\Sigma^{A}_{\phantom{A}A}=0$ and, antisymmetric (because $X{}_{A}{}^{F}{}_{C}{}_{F}=-X{}_{C}{}^{F}{}_{A}{}_{F}$). Hence $\Sigma_{AB}$ is already invariant under the action of the symmetric group and needs no further decomposition. 
Something similar happens to the last term of (\ref{eq:decompose-X-trace}).  

Thus it only remains to find the irreducible decomposition of $W_{ABCD}$. Since this tensor is totally traceless, we only need to work out its decomposition in parts invariant under the action of the symmetric group. The way in which this is achieved is by writing $W_{ABCD}$ as a sum of four rank tensors each of them being a {\em Young tableaux} tensor. A Young tableaux tensor is a tensor which is invariant under the action of a {\em Young projector} (see definition 5.6 of \cite{TUNG}). For any rank-4 tensor these are the Young tableaux which contribute to its decomposition 
\begin{eqnarray*}
&&\Yvcentermath1\young(A)\otimes\young(B)\otimes\young(C)\otimes\young(D)=
\young(ABCD)\oplus\young(AB,CD)\oplus\young(AC,BD)\oplus\young(ABC,D)
\oplus\young(ABD,C)\\
&&\Yvcentermath1\oplus\young(ACD,B)\oplus\young(AB,C,D)\oplus\young(AC,B,D)
\oplus\young(AD,B,C)\oplus\young(A,B,C,D).
\end{eqnarray*}
This decomposition is obtained by successive application of the {\em Littlewood-Richardson} rule to the product of Young tableaux which one obtains from the left hand side. If we now apply previous decomposition to $W_{ABCD}$ and take into account its symmetries we find that only the following Young tableaux contribute to the decomposition
\begin{equation}
\Yvcentermath1\young(ABCD)\;,\quad\young(AB,CD)\;,\quad\young(AC,BD).
\label{eq:young-tableaux}
\end{equation}
Therefore $W_{ABCD}$ is decomposed in the form
\begin{equation}
W{}_{A}{}_{B}{}_{C}{}_{D}=\Psi{}_{A}{}_{B}{}_{C}{}_{D}+
\Pi{}^{1}{}_{A}{}_{C}{}_{B}{}_{D} + 
\Pi{}^{2}{}_{A}{}_{B}{}_{C}{}_{D}.
\label{eq:decompose-W-step1}
\end{equation}
In this expression $\Psi_{ABCD}$ is a spinor with the symmetries of the first tableau in (\ref{eq:young-tableaux}), (i.e it is totally symmetric), and $\Pi{}^{1}{}_{A}{}_{C}{}_{B}{}_{D}$, $\Pi{}^{2}{}_{A}{}_{B}{}_{C}{}_{D}$ have the symmetries of the second and the third Young tableaux of (\ref{eq:young-tableaux}). These symmetries correspond to the unfilled tableau
$$
\yng(2,2)
$$ 
and therefore we deduce that both $\Pi^1_{ABCD}$ and $\Pi^2_{ABCD}$ fulfil the same algebraic properties as the Riemann tensor of a Levi-Civita connection, namely
\begin{eqnarray}
\Pi{}^{1}{}_{A}{}_{C}{}_{B}{}_{D}=-\Pi{}^{1}{}_{C}{}_{A}{}_{B}{}_{D}=
\Pi{}^{1}{}_{B}{}_{D}{}_{A}{}_{C}\;,\quad
\Pi{}^{1}{}_{A}{}_{C}{}_{B}{}_{D}+\Pi{}^{1}{}_{A}{}_{B}{}_{D}{}_{C}+
\Pi{}^{1}{}_{A}{}_{B}{}_{D}{}_{C}=0,\nonumber\\
\Pi{}^{2}{}_{A}{}_{C}{}_{B}{}_{D}=-\Pi{}^{2}{}_{C}{}_{A}{}_{B}{}_{D}=
\Pi{}^{2}{}_{B}{}_{D}{}_{A}{}_{C}\;,\quad
\Pi{}^{2}{}_{A}{}_{C}{}_{B}{}_{D}+\Pi{}^{2}{}_{A}{}_{B}{}_{D}{}_{C}+
\Pi{}^{2}{}_{A}{}_{B}{}_{D}{}_{C}=0.\nonumber\\
\label{eq:multiterm-symmetries}
\end{eqnarray}
The spinors $\Pi^1_{ABCD}$ and $\Pi^2_{ABCD}$ are not linearly independent. To see this take equation (\ref{eq:decompose-W-step1}) and antisymmetrize both sides of it on the indices $C,D$. The result is
$$
0=\frac{1}{2} \Pi {}^{1}{}_{A}{}_{C}{}_{B}{}_{D}-\frac{1}{2} \
\Pi {}^{1}{}_{A}{}_{D}{}_{B}{}_{C} + \Pi{}^{2}{}_{A}{}_{B}{}_{C}{}_{D}.
$$
Next, we use here the cyclic property $\Pi {}^{1}{}_{A}{}_{C}{}_{B}{}_{D}=
-\Pi{}^{1}{}_{A}{}_{B}{}_{D}{}_{C} -\Pi {}^{1}{}_{A}{}_{D}{}_{C}{}_{B}$ getting
$$
\Pi {}^{2}{}_{A}{}_{B}{}_{C}{}_{D}=-\frac{\Pi{}^{1}{}_{A}{}_{B}{}_{C}{}_{D}}{2}.
$$
If we replace this relation into (\ref{eq:decompose-W-step1}) we get
$$
W{}_{A}{}_{B}{}_{C}{}_{D}=-\frac{1}{2}\Pi{}^{1}{}_{A}{}_{B}{}_{C}{}_{D} + 
\Pi {}^{1}{}_{A}{}_{C}{}_{B}{}_{D}+ \Psi {}_{A}{}_{B}{}_{C}{}_{D}.
$$
We use here the property $\Pi {}^{1}{}_{A}{}_{B}{}_{C}{}_{D}=
-\Pi {}^{1}{}_{A}{}_{C}{}_{D}{}_{B}-\Pi {}^{1}{}_{A}{}_{D}{}_{B}{}_{C}$ which arises from the algebraic properties (\ref{eq:multiterm-symmetries}). This renders the decomposition of $W_{ABCD}$ in the form
\begin{equation}
W{}_{A}{}_{B}{}_{C}{}_{D}=\Psi {}_{A}{}_{B}{}_{C}{}_{D} + 
\Omega {}_{A}{}_{C}{}_{B}{}_{D} + \Omega {}_{A}{}_{D}{}_{B}{}_{C},
\label{eq:decompose-W-step2}
\end{equation}
where we have set
$$
\Omega_{ABCD}\equiv\frac{\Pi^1_{ABCD}}{2}.
$$
From this definition it is obvious that the spinor $\Omega_{ABCD}$ has the same algebraic properties as $\Pi^1_{ABCD}$, which are shown in (\ref{eq:multiterm-symmetries}).
Also from the tracelessness of $W_{ABCD}$ we deduce 
\begin{equation}
\Omega_{B\phantom{C}DC}^{\phantom{B}C}+\Omega_{BD\phantom{C}C}^{\phantom{BD}C}=0,
\label{eq:omega-traceless}
\end{equation}
from which, using the cyclic property $\Omega_{ABCD}=-\Omega_{ACDB}-\Omega_{ADBC}$ we obtain
\begin{equation}
 \Omega_{B\phantom{C}DC}^{\phantom{B}C}=0\;,\quad
\Omega_{BD\phantom{C}C}^{\phantom{BD}C}=0,
\label{eq:trace-of-omega}
\end{equation}
and therefore $\Omega_{ABCD}$ is completely traceless (as it should be since all non-vanishing trace-parts of $W_{ABCD}$ were taken away in the first step of the decomposition of $X_{ABCD}$).

We note that it is possible to obtain (\ref{eq:decompose-W-step2}) without any knowledge of Young tableaux theory if one takes the relations
$$
\Psi {}_{A}{}_{B}{}_{C}{}_{D}=\frac{1}{3}(
W{}_{A}{}_{B}{}_{C}{}_{D} +  W{}_{A}{}_{C}{}_{B}{}_{D} + \
 W{}_{A}{}_{D}{}_{B}{}_{C})\;,\quad
\Omega {}_{A}{}_{C}{}_{B}{}_{D}=\frac{1}{3}(
W{}_{A}{}_{B}{}_{C}{}_{D}-W{}_{A}{}_{D}{}_{B}{}_{C})
$$ 
as the definitions for the spinors $\Psi_{ABCD}$ and $\Omega_{ABCD}$ and then one deduces all the algebraic properties of these spinors straight from these definitions and the properties of $W_{ABCD}$.

Now, we substitute (\ref{eq:decompose-W-step2}) and (\ref{eq:decompose-X-step2}) into (\ref{eq:decompose-X-trace}) which yields the complete decomposition of $X_{ABCD}$
into irreducible parts
\begin{eqnarray}
X{}_{A}{}_{B}{}_{C}{}_{D}&=&\Lambda(\epsilon {}_{A}{}_{D} 
\epsilon{}_{B}{}_{C}+\epsilon{}_{A}{}_{C}\epsilon{}_{B}{}_{D})+\frac{1}{6}  (\epsilon {}_{B}{}_{D} \Sigma{}_{A}{}_{C} +\epsilon {}_{B}{}_{C} \Sigma {}_{A}{}_{D}\epsilon {}_{A}{}_{D} \Sigma {}_{B}{}_{C}+\epsilon {}_{A}{}_{C} \Sigma {}_{B}{}_{D})\nonumber\\  &&+\,\Psi{}_{A}{}_{B}{}_{C}{}_{D} + \Omega {}_{A}{}_{C}{}_{B}{}_{D} + \Omega{}_{A}{}_{D}{}_{B}{}_{C},
\label{eq:decompose-X-full}
\end{eqnarray}
where
$$
\Lambda\equiv\frac{X^{AB}_{\phantom{AB}AB}}{20}.
$$
At this stage we perform some basic counting to check the consistency of (\ref{eq:decompose-X-full}). On one hand we compute the number of independent components of the spinor $X_{ABCD}$ and on the other, we add the number of independent components of each of the irreducible parts in which $X_{ABCD}$ has been decomposed. The numbers are shown in table \ref{tab:independent-components}.

\begin{table}[h]
\begin{tabular}{|l|l|l|l|}
\hline
Quantity & symmetry-independent components & Restrictions & Final number \\ 
\hline
$\Lambda$ & 1 & 0 & 1\\ 
\hline
$\Omega_{ABCD}$ & 20 (Riemann-like rank-4 tensor in dim. 4) & 6 (use (\ref{eq:trace-of-omega})) & 14 \\
\hline
$\Psi_{ABCD}$ & 35 (totally symmetric rank-4 tensor) & 0 & 35\\
\hline
 $\Sigma_{AB}$  & 6 (antisymmetric, rank-2 tensor)  & 1
 ($\Sigma^{A}_{\phantom{A}A}=0$) & 5\\
\hline
$X_{ABCD}$    & 55 (use (\ref{eq:symmetries-of-X})) & 0 & 55\\
 \hline
\end{tabular}
\caption{In the first column of this table we show the spinor we deal with, in the second column we give the number of independent components of the spinor when only its symmetries are taken into account (together with a short explanation about how this number is computed), the third column shows the number of independent additional restrictions (if any) and their origin and the last column is just the difference of the symmetry-independent components and the restrictions. If we add all the entries of this column but the last one of the table we get 55 which is precisely the number of independent components of $X_{ABCD}$, thus confirming that (\ref{eq:decompose-X-full}) is indeed correct. \label{tab:independent-components}}
\end{table}
The number of independent components of the Riemann tensor in dimension 5 is 50 which means that one needs to impose additional restrictions on $X_{ABCD}$. These restrictions are just those arising from imposing the cyclic property $R_{abcd}+R_{bcad}+R_{cabd}=0$ on (\ref{eq:RiemannToXSpinor}) since the symmetries shown in (\ref{eq:symmetries-of-X}) only take into account the {\em monoterm symmetries} of the Riemann tensor. For our work we do not need to compute explicitly these additional relations and it is enough to realise that if we drop the spinor $\Sigma_{AB}$ from the decomposition (\ref{eq:decompose-X-full}) then the resulting quantity has precisely 50 independent components (indeed this is the only way of obtaining a quantity with 50 independent components out of (\ref{eq:decompose-X-full}), see table \ref{tab:independent-components}). Inserting the irreducible decomposition of $X_{ABCD}$ into (\ref{eq:RiemannToXSpinor}) with $\Sigma_{AB}$ set to zero and summarising the results found before we can state 
the following result.
\begin{theorem}
The Riemann tensor $R_{abcf}$ of the covariant derivative $\nabla_a$ can be decomposed in the form
\begin{equation}
R{}_{a}{}_{b}{}_{c}{}_{f}=\Lambda(g{}_{a}{}_{f} g{}_{b}{}_{c}  
-g{}_{a}{}_{c} g{}_{b}{}_{f})-\frac{1}{2} 
G{}_{a}{}_{b}{}^{A}{}^{B} G{}_{c}{}_{f}{}^{C}{}^{D}
\Psi{}_{A}{}_{B}{}_{C}{}_{D}-G{}_{a}{}_{b}{}^{A}{}^{B}
G{}_{c}{}_{f}{}^{C}{}^{D}\Omega {}_{A}{}_{C}{}_{B}{}_{D}.
\label{eq:decompose-Riemann-full}
\end{equation}
The quantities $\Lambda$, $\Omega_{ABCD}$ and $\Psi_{ABCD}$ are known collectively as the \underline{curvature spinors}.  Furthermore, the curvature spinors fulfil the algebraic properties
\begin{eqnarray}
&&\Psi_{(ABCD)}=\Psi_{ABCD}\;,\quad
\Omega_{ABCD}=\Omega_{[AB]CD}=\Omega_{CDAB}\;,\quad
\Omega_{AB\phantom{C}C}^{\phantom{AB}C}=\Omega_{A\phantom{C}CD}^{\phantom{A}C}=0\;,
\nonumber\\
&&\Omega_{ABCD}+\Omega_{BCAD}+\Omega_{CABD}=0.
\label{eq:properties-of-curvature-spinors}
\end{eqnarray}
\label{theorem:curvature-spinors}
\end{theorem}
\qed
\begin{remark}\em
 Taking traces in the formula (\ref{eq:decompose-Riemann-full}), using 
(\ref{eq:G-definition}) and (\ref{eq:Gsquares}) we can obtain the 
decomposition of the Ricci tensor and the scalar curvature. The actual expressions are
\begin{equation}
R{}_{a}{}_{c}=-4 g{}_{a}{}_{c} \Lambda-\frac{1}{2}\gamma{}_{a}{}^{A}{}^{B}
\gamma {}_{c}{}^{C}{}^{D} 
(\Omega{}_{A}{}_{B}{}_{C}{}_{D} + \Omega {}_{A}{}_{C}{}_{B}{}_{D})\;,\quad 
R=-20\Lambda. 
\label{eq:ricci-ricci-scalar}
\end{equation}
Previous expression adopts a simpler form if we apply the identity
$$
\gamma {}_{a}{}^{A}{}^{B} \gamma {}_{c}{}^{C}{}^{D}(\Omega{}_{A}{}_{B}{}_{C}{}_{D} +\Omega{}_{A}{}_{C}{}_{B}{}_{D})=\frac{3}{2}\gamma {}_{a}{}^{A}{}^{B}
\gamma {}_{c}{}^{C}{}^{D}\Omega {}_{A}{}_{B}{}_{C}{}_{D}.
$$
To obtain it we replace $\Omega_{ACBD}$  by $(\Omega_{ACBD}-\Omega_{BCAD})/2$ in the left hand side and then use the replacement $\Omega_{ACBD}=-\Omega_{ABDC}-\Omega_{ADCB}$ on the resulting expression. In fact, it is more convenient to write (\ref{eq:ricci-ricci-scalar}) in terms of the {\em traceless Ricci tensor}, $S_{ab}\equiv R_{ab}-R g_{ab}/5$
\begin{equation}
S{}_{a}{}_{c}=-\frac{3}{4}\gamma{}_{a}{}^{A}{}^{B}
\gamma {}_{c}{}^{C}{}^{D}\Omega{}_{A}{}_{B}{}_{C}{}_{D}.
\label{eq:traceless-ricci-tensor}
\end{equation}
Previous equation expresses the fact that $\Omega_{ABCD}$ is indeed the spinor counterpart of the traceless Ricci tensor. 
We follow the 4-dimensional terminology and refer to 
$\Omega_{ABCD}$ as the ``Ricci spinor'' although the suitable name would be the ``traceless Ricci spinor''.
\end{remark}
The scalar $\Lambda$ and the Ricci spinor vanish if and only if $S_{ab}=0$, $R=0$ in which case $R_{abcd}$ becomes the Weyl tensor $C_{abcd}$. Therefore by setting $\Omega_{ABCD}=0$, $\Lambda=0$ on (\ref{eq:decompose-Riemann-full}) we deduce 
\begin{equation}
C_{abcf}=-\frac{1}{2} 
G{}_{a}{}_{b}{}^{A}{}^{B} G{}_{c}{}_{f}{}^{C}{}^{D}
\Psi{}_{A}{}_{B}{}_{C}{}_{D} 
\label{eq:weyl-spinor}
\end{equation}
and hence the spinor $\Psi_{ABCD}$ has all the information about the Weyl tensor. Again we use the 4-dimensional nomenclature and call $\Psi_{ABCD}$ the Weyl spinor. 
Note that (\ref{eq:weyl-spinor}) is still true when the Ricci spinor and $\Lambda$ do not vanish because the Weyl tensor is linearly independent from any quantity containing the Ricci scalar and the trace-free Ricci tensor.  
Equation (\ref{eq:weyl-spinor}) has been already presented in \cite{SMET1} where the Weyl spinor was used to build an algebraic classification of the Weyl tensor $C_{abcd}$. 

Equation (\ref{eq:decompose-Riemann-full}) bears a strong resemblance with the formula which yields the curvature spinors in 4-dimensional Lorentzian geometry. In that case one has three types of curvature spinors and two of them are of rank four as well (the Weyl spinor and the Ricci spinor). In the case of 4-dimensional Lorentzian geometry all the curvature spinors have mono-term symmetries but this is not true in the 5-dimensional case where the Ricci spinor $\Omega_{ABCD}$ fulfils the {\em cyclic property}. As far as we know this is the first time in which the general decomposition of the Riemann tensor (\ref{eq:decompose-Riemann-full}) in a 5-dimensional spacetime is computed. 
 
\subsection{Spinor form of the second Bianchi identity}
As in the case of the spinor calculus of 4-dimensional Lorentzian geometry, the curvature spinors introduced above fulfil a differential identity which is equivalent to the second Bianchi identity of the Riemann tensor. We present this identity in the next proposition and we explain how it is obtained.
\begin{prop}
 The curvature spinors $\Psi_{ABCD}$ and $\Omega_{ABCD}$ satisfy the following
differential identity
\begin{equation}
\nabla_{(Z}{}^{W}\Psi {}_{V)}{}_{B}{}_{A}{}_{W}-
\nabla_{(Z}{}^{W}\Omega {}_{V)}{}_{A}{}_{B}{}_{W}- 
\nabla_{(Z}{}^{W}\Omega {}_{V)BAW}-
2\epsilon{}_{(A}{}_{|(V}\nabla_{Z)|}{}_{B)}\Lambda=0. 
\label{eq:spinor-bianchi-identity}
\end{equation}
\label{prop:bianchi-identity}
\end{prop}

\noindent
\Proof
We start with the Bianchi identity (\ref{eq:F-bianchi-identity}) particularised to the covariant derivative $\nabla_a$ and we replace the inner curvature by its formula in terms of $X_{ABCD}$ given in (\ref{eq:FRiemanToX}). This gives
\begin{equation}
G{}_{b}{}_{c}{}_{C}{}_{D} 
\nabla{}_{a}X{}^{C}{}^{D}{}_{A}{}_{B} + G{}_{c}{}_{a}{}_{C}{}_{D} 
\nabla{}_{b}X{}^{C}{}^{D}{}_{A}{}_{B} + G{}_{a}{}_{b}{}_{C}{}_{D} 
\nabla{}_{c}X{}^{C}{}^{D}{}_{A}{}_{B}=0.
\label{eq:second-bianchi-X}
\end{equation}
We replace here $G_{abAB}$ using (\ref{eq:G-definition}) getting
$$
-\gamma {}_{b}{}^{C}{}^{D}\gamma {}_{c}{}_{C}{}^{F} 
\nabla{}_{a}X{}_{A}{}_{B}{}_{D}{}_{F} + \gamma {}_{a}{}^{C}{}^{D} \
\gamma {}_{c}{}_{C}{}^{F} 
\nabla{}_{b}X{}_{A}{}_{B}{}_{D}{}_{F}-\gamma {}_{a}{}^{C}{}^{D}
\gamma {}_{b}{}_{C}{}^{F}\nabla{}_{c}X{}_{A}{}_{B}{}_{D}{}_{F}=0.
$$ 
The quantities $\gamma$ 
in this expression can be removed if we multiply both sides of it by 
$\gamma {}^{a}{}_{Y}{}_{W}\gamma {}^{b}{}_{U}{}_{V}\gamma{}^{c}{}_{T}{}_{Z}$
and use (\ref{eq:gammasquare}) where necessary. The final expression is a bit 
long but it can be shortened if we contract the free index $T$ with $W$ and the free index $U$ with $Y$. The result of these operations is
\begin{equation}
\nabla_{(Z}{}^{W}X{}_{V)}{}_{W}{}_{A}{}_{B}=0.
\label{eq:bianchi-identity-X}
\end{equation}
Note that this expression has the same information as (\ref{eq:second-bianchi-X}) as can be checked for instance by counting the independent number of equations supplied by each of them when they are written in a generic frame (this number is 100). 
Now, it only remains to insert the decomposition (\ref{eq:decompose-X-full}) of the spinor $X_{ABCD}$ into (\ref{eq:bianchi-identity-X}). After doing this and going through some algebra (\ref{eq:spinor-bianchi-identity}) is finally derived.
\qed

\begin{remark}\em
 Given the relation (\ref{eq:RiemannToFRiemann}), the bianchi identity for the inner 
curvature $F_{abAB}$ is equivalent to the second Bianchi identity of the Riemann tensor $R_{abcd}$. Thus (\ref{eq:spinor-bianchi-identity}) has the same information as the second Bianchi identity of the Riemann tensor. 
\end{remark}

Equation (\ref{eq:spinor-bianchi-identity}) has a strong resemblance to the
differential identity fulfiled by the curvature spinors in the spinor calculus of 4-dimensional Lorentzian geometry. 

\subsection{The Ricci identity}
Consider now the Ricci identity (\ref{eq:ricci-identity-F}) particularised for $\nabla_a$. In order to handle this identity in an easier way, we define the linear operator
\begin{equation}
\DAl_{AB}\equiv G^{ab}_{\phantom{ab}AB}\nabla_a\nabla_b.
\label{eq:daloperator} 
\end{equation}
Straightforward properties of the operator $\DAl_{AB}$ are
\begin{equation}
\DAl_{(AB)}=\DAl_{AB}\;,\quad\DAl_{AB}(\xi_{A}\chi_B)=\chi_B\DAl_{AB}\xi_A+\xi_A\DAl_{AB}\chi_B. 
\label{eq:DAl-leibnitz}
\end{equation}
The Leibnitz rule is easily generalised to the product of two spinors of arbitrary rank. Next we find the value of the action of $\DAl_{AB}$ on any rank-1 spinor $\xi^A$. To that end we take (\ref{eq:ricci-identity-F}) and multiply both sides of it by $G^{ab}_{\phantom{ab}CD}$. The result is
$$
\DAl{}_{C}{}_{D}\xi^{B}=-\frac{1}{2}F{}_{a}{}_{b}{}_{A}{}^{B} \
G{}^{a}{}^{b}{}_{C}{}_{D} \xi^{A}.
$$  
We use in this formula the relation (\ref{eq:FRiemanToX}), getting
$$
\DAl{}_{C}{}_{D} 
\xi{}^{B}=-\frac{1}{4} X{}^{F}{}^{H}{}_{A}{}^{B} \
\xi{}^{A} (\epsilon {}_{C}{}_{H}\epsilon {}_{D}{}_{F} + \epsilon{}_{C}{}_{F}\epsilon {}_{D}{}_{H}),
$$
where the first of (\ref{eq:Gsquares}) was used along the way. Finally we apply the decomposition of $X{}^{F}{}^{H}{}_{A}{}^{B}$ shown in (\ref{eq:decompose-X-full})
obtaining (we lower all indices)
\begin{equation}
\DAl_{C}{}_{D}\xi{}_{B}=
\Lambda \epsilon {}_{B}{}_{(C}\xi_{D)}-\xi{}^{A} \Omega{}_{B}{}_{(C}{}_{D)}{}_{A} -\frac{1}{2}\xi{}^{A}\Psi{}_{B}{}_{C}{}_{D}{}_{A}.
\label{eq:DAl-action}  
\end{equation}
Using the Leibnitz rule (\ref{eq:DAl-leibnitz}) we can extend previous result to a spinor of arbitrary rank. We note the similarity of (\ref{eq:DAl-action}) with the action of the operator which is introduced in the spinor calculus of 4-dimensional Lorentzian geometry when studying the spinor form of the Ricci identity (\ref{eq:ricci-identity-v}). This operator has similar properties as the $\DAl_{AB}$ studied here and that is the reason why we chose the same notation for it as in the 4-dimensional case. 

\section{The extension of the Newman-Penrose formalism}
\label{sec:newman-penrose}
A very important application of the spinor calculus in 4-dimensional
Lorentzian geometry is the Newman-Penrose formalism \cite{NP}. The Newman-Penrose formalism consists in writing out all the Ricci and Bianchi identities in a null tetrad. By doing so one is able to set up a direct link between the components of the 
curvature spinors and the components of the Riemann tensor. Also the Ricci rotation coefficients can be written in terms of the spin coefficients which are complex quantities. This enables one to reduce the number of equations when we regard equations which are complex conjugate of each other as dependent.

Using the ideas developed in the foregoing sections we can achieve an {\em extension} of the Newman-Penrose formalism to a 5-dimensional spacetime. The word ``extension'' is appropriate here because our formalism contains the Newman-Penrose formalism as a particular case (which is reasonable given that a 4-dimensional spacetime is in some sense a ``subset'' of a 5-dimensional one). This means that all the variables which appear in the Newman-Penrose formalism (spin coefficients, curvature components, etc) will be also present in our equations.    

In this section we present the basic variables which we use in our extension
of the Newman-Penrose formalism together with some of the 5-dimensional Newman-Penrose equations. The full set of equations will be shown and studied elsewhere \cite{AGPLODE}.
 

\subsection{Components of the curvature spinors in a spin tetrad}
\label{subsection:curvature-spinors-components}
The first step is the introduction of suitable symbols for each of the independent components of the curvature spinors with respect to a spin tetrad, much in the way as it is done in the case of 4-dimensional Lorentzian geometry. This is done next separately for each curvature spinor.

These are the sixteen complex independent components of the Weyl spinor
\begin{equation}
\begin{array}{ccc}
\Psi_0\equiv\Psi_{ABCD}o^A o^B o^C o^D, & 
\ ^*\Psi_0\equiv\Psi_{ABCD}o^A o^B o^C \tilde o^D,  &   \\
\Psi_1\equiv\Psi_{ABCD}o^A o^B o^C \iota^D, &
\ ^*\Psi_1\equiv\Psi_{ABCD}o^A o^B \iota^C \tilde o^D,  &
\Psi^*_1\equiv\Psi_{ABCD}o^A o^B o^C \tilde\iota^D, \\
\Psi_2\equiv\Psi_{ABCD}o^A o^B \iota^C \iota^D, &
\ ^*\Psi_{2}\equiv\Psi_{ABCD}o^A\iota^B \iota^C \tilde o^D,  & \Psi^*_2\equiv\Psi_{ABCD}o^A o^B \iota^C \tilde\iota^D, \\
\Psi_3\equiv\Psi_{ABCD}o^A \iota^B \iota^C \iota^D, &
\ ^*\Psi_{3}\equiv\Psi_{ABCD}\tilde o^A\iota^B\iota^C \iota^D,  & \Psi^*_{3}\equiv\Psi_{ABCD}o^A\iota^B \iota^C \tilde\iota^D, \\
\Psi_4\equiv\Psi_{ABCD}\iota^A \iota^B \iota^C \iota^D, & \Psi^*_{4}\equiv\Psi_{ABCD}\iota^A\iota^B\iota^C \tilde\iota^D,
 & \\
\Psi_{01}\equiv\Psi_{ABCD}o^A o^B \tilde o^C \tilde\iota^D, &\Psi_{02}\equiv\Psi_{ABCD}o^A o^B \tilde\iota^C \tilde\iota^D & 
\Psi_{12}\equiv\Psi_{ABCD}o^A \iota^B \tilde\iota^C \tilde\iota^D.\\
\end{array}
\label{eq:weyl-spinor-complex-components}
\end{equation}
The sub-index in the scalars $\Psi_0$-$\Psi_4$ indicates the number of $\iota$ spinors used in the contraction with the Weyl spinor (this notation is similar to the 4-dimensional case and indeed these scalars correspond to the components of the 4-dimensional Weyl spinor when the 4-dimensional reduction is performed, see \S\ref{subsubsec:4-dim-reduction}). We use these scalars as the starting point to denote other components of the Weyl spinor as follows:  a star in front of a scalar means that the replacement $o^A\rightarrow \tilde{o}^A$ has been performed \underline{once} in the definition which yields the scalar whereas a star behind a scalar means that the replacement $\iota^A\rightarrow \tilde{\iota}^A$
is done (also once). For example starting from $\Psi_2$ we define $\ ^*\Psi_2$ (resp. $\Psi^*_2$) as follows (see (\ref{eq:weyl-spinor-complex-components}))
$$
\Psi_2=\Psi_{ABCD}o^A o^B \iota^C \iota^D\Rightarrow
\ ^*\Psi_2\stackrel{o^A\rightarrow \tilde{o}^A}{=}
\Psi_{ABCD}\tilde{o}^A o^B \iota^C \iota^D
\Rightarrow
\Psi^*_2\stackrel{\iota^C\rightarrow \tilde{\iota}^C}{=}\Psi_{ABCD}o^A o^B \tilde{\iota}^C \iota^D.
$$
When we have a scalar with the same number of elements of the spin tetrad with no tilde as elements with tilde we append two sub-indices to the scalar.
Each of these sub-indices tells, respectively, the number of spinors $\iota$ and $\tilde{\iota}$ intervening in the definition of this scalar (see the definitions of 
$\Psi_{01}$, $\Psi_{02}$, $\Psi_{12}$).  The notation conventions just introduced for these components of the Weyl spinor are kept for the remaining components of the curvature spinors. Given the symmetries of $\Psi_{ABCD}$ and $\Omega_{ABCD}$ these conventions should lead to no confusion.

These are the three real components of the Weyl spinor.
\begin{equation}
\Psi_{00}\equiv\Psi_{ABCD}o^A o^B \tilde o^C \tilde o^D\;,\quad
\Psi_{11}\equiv\Psi_{ABCD}o^A \iota^B \tilde o^C \tilde \iota^D\;,\quad
\Psi_{22}\equiv\Psi_{ABCD}\iota^A \iota^B \tilde\iota^C \tilde\iota^D\;,\quad
\label{eq:weyl-spinor-real-components}
\end{equation}
These are the four complex components of the Ricci spinor
\begin{eqnarray}
&&\Phi_{01}\equiv\Omega_{ABCD} o^A\tilde o^B o^C \tilde\iota^D\;,\quad
\Phi_{02}\equiv\Omega_{ABCD} o^A\tilde \iota^B o^C \tilde\iota^D\;,\quad
\Phi_{12}\equiv\Omega_{ABCD} o^A\tilde\iota^B\iota^C\tilde\iota^D\;,\quad\nonumber\\
&&\ ^*\Phi_{02}\equiv\Omega_{ABCD} o^A\tilde\iota^B\tilde o^C\tilde\iota^D.
\label{eq:ricci-spinor-complex-components}
\end{eqnarray}
These are the six real components of the Ricci spinor.
\begin{eqnarray}
&&\Phi_{00}\equiv\Omega_{ABCD}o^A\tilde o^B o^C \tilde o^D\;,\quad
\Phi_{11}\equiv\Omega_{ABCD}o^A\tilde o^B \iota^C \tilde\iota^D\;,\quad
\Phi_{22}\equiv\Omega_{ABCD}\iota^A\tilde\iota^B \iota^C \tilde\iota^D\;,\quad\nonumber\\
&&\Omega\equiv\Omega_{ABCD}\tilde o^A\tilde\iota^B \tilde o^C \tilde\iota^D\;,\quad
\ ^*\Phi_{01}\equiv\Omega_{ABCD}o^A\tilde o^B \tilde o^C \tilde\iota^D\;,\quad
\ ^*\Phi_{12}\equiv\Omega_{ABCD}\iota^A\tilde\iota^B \tilde o^C \tilde\iota^D\;,\quad
\label{eq:ricci-spinor-real-components}
\end{eqnarray}
To the quantities introduced in the previous paragraphs, one should add the scalar curvature $\Lambda$.
A simple counting shows that the number of these quantities and the number of their complex conjugates (when they are complex) adds up to 50, which is the number of independent components of the Riemann tensor. Also these quantities contain as a subset the Newman-Penrose scalars which are used to represent the components of the Riemann tensor (these are denoted with the usual symbols used within the Newman-Penrose formalism so the reader can spot them).  

\subsection{The commutation relations}
The commutation relations of the operators $\nabla_{\bf 1}=D$, 
$\nabla_{\bf 2}=\Delta$, $\nabla_{\bf 3}=\delta$, $\nabla_{\bf 4}=\bar\delta$, $\nabla_{\bf 5}=\mathcal D$ are given by the equation
$$
\nabla{}_{{\mathbf a}}\nabla{}_{{\mathbf b}}Z -\nabla{}_{\mathbf b}\nabla{}_{\mathbf a}Z
=T{}^\mathbf{1}{}_{\mathbf b}{}_{\mathbf a}D Z+T{}^\mathbf{2}{}_{\mathbf b}{}_{\mathbf a}\Delta Z+T{}^\mathbf{3}{}_{\mathbf b}{}_{\mathbf a}\delta Z+
T{}^\mathbf{4}{}_{\mathbf b}{}_{\mathbf a}\bar\delta Z+
T{}^\mathbf{5}{}_{\mathbf b}{}_{\mathbf a}{\mathcal D} Z,
$$   
where $Z$ is an arbitrary scalar field and 
\begin{equation}
T^{\mathbf a}_{\phantom{\mathbf a}{\mathbf b}{\mathbf c}}=
\Gamma(\nabla,\mathcal N)^{\mathbf a}_{\phantom{\mathbf a}{\mathbf c}{\mathbf b}}-
\Gamma(\nabla,\mathcal N)^{\mathbf a}_{\phantom{\mathbf a}{\mathbf b}{\mathbf c}}.
\label{Torsion}
\end{equation}
All the quantities 
$\Gamma(\nabla,\mathcal N)^{\mathbf a}_{\phantom{\mathbf a}{\mathbf c}{\mathbf b}}$
can be written in terms of the spin coefficients by means of (\ref{eq:riccirotation-to-spin-coeffs}).
Making the appropriate replacements we get (we suppress the arbitrary scalar $Z$)
\begin{eqnarray}
D\Delta-\Delta D &=&
-(\gamma + \bar\gamma)D -(\epsilon + \bar\epsilon)\Delta 
-(\pi  + \bar\tau)\delta-(\bar\pi + \tau )\bar\delta- 
(\mathfrak{a} + \mathfrak{e}){\mathcal D}
\;,\\
D\delta-\delta D &=&
-(\bar\alpha + \beta  + \bar\pi)D + \kappa  \Delta + 
(\epsilon-\bar\epsilon-\bar\rho)\delta
-\sigma\bar\delta -(\varsigma -\chi ){\mathcal D}
\;,\\
D{\mathcal D}-{\mathcal D}D &=&
(-2\mathfrak{a} + \theta  + \bar\theta)D + 2 \mathfrak{d} \Delta
+ (\bar\eta -2 \bar\chi)\delta + (\eta  -2 \chi )\bar\delta
- \mathfrak{f} {\mathcal D}
\;,\\
\Delta\delta-\delta\Delta &=&
-\bar\nu D  + (\bar\alpha + \beta  + \tau )\Delta
+ (\gamma -\bar\gamma + \mu )\delta  + \bar\lambda \bar\delta
+(\xi  + \omega ){\mathcal D}
\;, \\
\Delta{\mathcal D}-{\mathcal D}\Delta&=&
-2 \mathfrak{b} D + (2 \mathfrak{e} -\theta -\bar\theta)\Delta 
+ (\zeta  -2\bar\omega)\delta
+ (\bar\zeta -2 \omega )\bar\delta +\mathfrak{c} {\mathcal D}
\;,\\ 
\delta\bar\delta-\bar\delta\delta &=&
(-\mu +\bar\mu)D + (-\rho  + \bar\rho)\Delta  +
(-\alpha  +\bar\beta)\delta  + (\bar\alpha-\beta )\bar\delta -   
(\upsilon-\bar\upsilon){\mathcal D}
\;,\\
\delta{\mathcal D}-{\mathcal D}\delta &=&
(\bar\zeta-2\xi )D + (\eta  + 2 \varsigma )\Delta   +
(\theta -\bar\theta -2\bar\upsilon)\delta   -2 \phi  \bar\delta  
- \psi  {\mathcal D}
\;. 
\end{eqnarray}
These equations reduce to the standard Newman-Penrose commutation relations when we set the 5-dimensional spin coefficients and $\mathcal D$ to zero. 

\subsection{The Components of the Riemann tensor}
In this subsection we show the relation between the components of the Riemann tensor in a semi-null pentad and the quantities introduced in \S\ref{subsection:curvature-spinors-components}. The starting point is equation (\ref{eq:decompose-Riemann-full}) which is expressed in the spin tetrad and the semi-null pentad. The components in these frames of $G_{ab}^{\phantom{ab}AB}$ can be readily computed using (\ref{eq:G-definition}) and (\ref{eq:gamma-expansion}) so we only need to insert the corresponding values. After some algebra, the independent components of the Riemann tensor turn out to be 
\begin{eqnarray} &&
R_\mathbf{1212}=2 \Phi _{11} + \Lambda -\frac{1}{2}(\Psi _2 +\overline{\Psi} _2{})+\Psi_{11}\;, \
R_\mathbf{1213}=-\Phi _{01} + \frac{1}{2}(\Psi _1-\Psi_{01})\;,
\nonumber\\ &&
R_\mathbf{1215}=\ ^*\Psi_1 +\ ^*\overline{\Psi}_1{} -2 \ ^*\Phi_{01}\;,\
R_\mathbf{1223}=\Phi_{12}-\frac{1}{2}(-\Psi_{12} +\overline{\Psi}_3{})\;,
\nonumber\\ &&
R_\mathbf{1225}=\Psi^*_3+\overline{\Psi}^*_3 + 2\ ^*\Phi_{12}\;,\
R_\mathbf{1234}=\frac{1}{2}(\Psi _2 -\overline{\Psi}_2)\;,
\nonumber\\ &&
R_\mathbf{1235}=-\ ^*\overline{\Psi}_{2} - \Psi^*_2\;,\
R_\mathbf{1313}=-\frac{1}{2} \Psi_0\;,\  
R_\mathbf{1314}=\Phi _{00}+\frac{1}{2} \Psi_{00}\;,\
R_\mathbf{1315}=-\ ^*\Psi_0\;,
\nonumber\\ &&
R_\mathbf{1323}=-\Phi _{02} - \frac{1}{2} \Psi_{02}\;, \
R_\mathbf{1324}=\Lambda + \frac{1}{2} \Psi _2+\Omega\;,\
R_\mathbf{1325}=-\Psi^*_2 -2 \ ^*\Phi_{02}\;,
\nonumber\\ &&
R_\mathbf{1334}=-\Phi _{01}-\frac{1}{2} (\Psi _1 +\Psi_{01})\;,\
R_\mathbf{1335}=\Psi^*_1\;, \
R_\mathbf{1345}=\ ^*\Psi_1 + 2 \ ^*\Phi_{01}\;,
\nonumber\\ &&
R_\mathbf{1515}=2 (\Phi _{00} -\Psi _{00})\;, \
R_\mathbf{1523}=\ ^*\overline{\Psi}_2 -2 \ ^*\Phi_{02}\;, \
R_\mathbf{1525}=2(\Phi _{11} +\Lambda -\Psi_{11} - 2\Omega)\;,
\nonumber\\ &&
R_\mathbf{1534}=-\ ^*\Psi_1 +\ ^*\overline{\Psi}_1\;, \
R_\mathbf{1535}=-2(\Phi _{01} -\Psi_{01})\;, \
R_\mathbf{2323}=-\frac{1}{2} \overline{\Psi}_4\;,
\nonumber\\ &&
R_\mathbf{2324}=\Phi _{22} +\frac{1}{2} \Psi_{22}\;, \
R_\mathbf{2325}=\overline{\Psi}^*_4\;, \
R_\mathbf{2334}=-\Phi _{12} + \frac{1}{2}(-\Psi_{12}-\overline{\Psi}_3)\;,
\nonumber\\ &&
R_\mathbf{2335}=-\ ^*\overline{\Psi}_3\;, \
R_\mathbf{2345}=-\overline{\Psi}^*_3 + 2 \ ^*\Phi_{12}\;, \
R_\mathbf{2525}=2(\Phi _{22} -\Psi_{22})\;,
\nonumber\\ &&
R_\mathbf{2534}=-\Psi^*_3 +\overline{\Psi}^*_3\;, \
R_\mathbf{2535}=-2(\Phi _{12}-\Psi_{12})\;,
\nonumber\\ &&
R_\mathbf{3434}=2 \Phi _{11} + \Lambda - \frac{1}{2}(\Psi _2 +\overline{\Psi} _2)- \Psi_{11} + 2 \Omega\;, \
R_\mathbf{3435}=-\ ^*\overline{\Psi}_2+\Psi^*_2 -2 \ ^*\Phi_{02}\;,
\nonumber\\ &&
R_\mathbf{3535}=2(\Phi _{02} -\Psi_{02})\;, \
R_\mathbf{3545}=2(\Phi _{11} - \Lambda -\Psi_{11} +\Omega)\;.
\label{eq:riemann-tensor-components} 
\end{eqnarray}
This is a set of 34 equations which has all the information about the Riemann tensor.
Now, one can write the Riemann tensor components in terms of the Ricci rotation coefficients by means of the standard formula
\begin{eqnarray}
R{}_{\mathbf a}{}_{\mathbf b}{}_{\mathbf c}{}_{\mathbf d}&=&
g{}_{\mathbf d}{}_{\mathbf f}
\Big(\Gamma(\nabla,{\mathcal N}){}^{\mathbf f}{}_{\mathbf b}{}_{\mathbf h} \Gamma(\nabla,\mathcal{N}){}^{\mathbf h}{}_{\mathbf a}{}_{\mathbf c}-
\Gamma(\nabla,\mathcal{N}){}^{\mathbf f}{}_{\mathbf a}{}_{\mathbf h}
\Gamma(\nabla,\mathcal{N}){}^{\mathbf h}{}_{\mathbf b}{}_{\mathbf c}
\nonumber\\ && \qquad
-\Gamma(\nabla,\mathcal{N}){}^{\mathbf f}{}_{\mathbf h}{}_{\mathbf c}
T{}^{\mathbf h}{}_{\mathbf a}{}_{\mathbf b}
-\nabla{}_{\mathbf a}
\Gamma(\nabla,\mathcal{N}){}^{\mathbf f}{}_{\mathbf b}{}_{\mathbf c} +
\nabla{}_{\mathbf b}
\Gamma(\nabla,\mathcal{N}){}^{\mathbf f}{}_{\mathbf a}{}_{\mathbf c}\Big).
\label{eq:riemann-to-riccirotation}
\end{eqnarray}
If we replace here the values of the Riemann tensor components found in (\ref{eq:riemann-tensor-components}) and the Ricci rotation coefficients by their values in terms of the spin coefficients, shown in (\ref{eq:riccirotation-to-spin-coeffs}), we get a set of 34 equations which can be used as the starting point to extend the 4-dimensional Newman-Penrose equations to a 5-dimensional spacetime. The complete set of this equations shall be presented an analysed in \cite{AGPLODE}. 

\subsubsection{4-dimensional reduction}
\label{subsubsec:4-dim-reduction}
It is instructive to study how the standard 4-dimensional Newman-Penrose formalism is recovered
from the 5-dimensional equations (one does not need to know the full set of 5-dimensional equations in order to study this reduction). To perform the reduction first we set to zero the spin coefficients defined by (\ref{eq:define-complexcoefficients})-(\ref{eq:define-realcoefficients}). Next we must compute the conditions on the components of the curvature spinors which lead to the reduction. These conditions are given by the relations  
\begin{eqnarray*}
&& R_\mathbf{1215}=R_\mathbf{1225}=R_\mathbf{1235}=R_\mathbf{1315}=R_\mathbf{1325}=R_\mathbf{1335}=R_\mathbf{1345}=R_\mathbf{1515}=\\
&&=R_\mathbf{1523}=R_\mathbf{1525}=R_\mathbf{1534}=R_\mathbf{1535}=R_\mathbf{2325}=R_\mathbf{2335}=R_\mathbf{2345}=R_\mathbf{2525}=\\
&&=R_\mathbf{2534}=R_\mathbf{2535}=R_\mathbf{3435}=R_\mathbf{3535}=R_\mathbf{3545}=0,
\end{eqnarray*}
which mean that there is a local coordinate system $\{x^1,x^2,x^3,x^4,x^5\}$ on ${\mathcal M}$ such that $u^a=\partial/\partial x^5$ on it and the metric tensor takes the form
$$
ds^2=-(dx^5)^2+\sum^4_{i,j=1}g_{ij}dx^idx^j,
$$ 
with the functions $g_{ij}$ only depending on $(x^1,x^2,x^3,x^4)$, i.e. the metric tensor is {\em decomposable} or {\em reducible} \cite{AGP}.
If we replace here the values of the components of the Riemann tensor found in (\ref{eq:riemann-tensor-components}) we get
\begin{eqnarray}
&&\Psi_{02}=\Phi _{02}\;,\quad \ ^*\Psi_2=0\;,\quad \Psi^*_3=0\;,\quad 
\Psi_{13}=\Phi _{12}\;,\quad \ ^*\Psi_3=0\;,\quad \Psi^*_4=0\;,\nonumber\\ 
&&\ ^*\Psi_0=0\;,\quad \
\Psi^*_1=0\;,\quad \ ^*\Psi_1=0\;,\quad \Psi^*_2=0\;,\quad \Psi_{01}=\Phi _{01}\;,\quad \
\Psi_{00}=\Phi _{00}\;,\nonumber\\ 
&&\hspace{-.5cm}\Psi_{11}= \Phi _{11} -\frac{\Lambda}{3}\;,\quad \Psi_{22}=\Phi _{22}\;,\quad 
\ ^*\Phi_{02}=0\;,\quad \Omega=\frac{2\Lambda}{3}  \;,\quad \ ^*\Phi_{01}=0 \;,\quad \ ^*\Phi_{12}=0.
\label{eq:4Dcurvature-reduction} 
\end{eqnarray}
From this expression we deduce that some components of the 4-dimensional Ricci-spinor  are related to components of the 5-dimensional Weyl spinor.  

\section{Conclusions}
\label{sec:conclusions}
We have studied in detail the spinor calculus in a 5-dimensional Lorentzian manifold  and show how concepts so important as the spin structure, spin covariant derivative and the curvature spinors are defined in this framework. The algebraic and differential properties of the curvature spinors have been studied in detail. An interesting application of the ideas presented in this paper is the extension of the Newman-Penrose formalism to a 5-dimensional spacetime. In this regard we have shown how by using the spinor techniques one can define a set of quantities (the spin coefficients and the components of the curvature spinors in a spin tetrad) which contains the variables used in the 4-dimensional Newman-Penrose formalism. This means that when certain quantities in our formalism are set to zero one recovers the usual 4-dimensional Newman-Penrose formalism (this corresponds to a {\em dimensional reduction} from five to four dimensions). In particular we have seen that under this dimensional reduction some components of the 4-dimensional Ricci spinor are directely related to components of the 5-dimensional Weyl spinor. 
Also the introduction of complex spin coefficients and the curvature spinors enables us to reduce the number of variables in our frame formalism. For example we only need to deal with 28 spin coefficients,
and 30 curvature scalars, rather than the 50 quantities which we would need in each case if we used a frame formalism not based on the spinor approach.  

We have given the explicit form of the commutation relations of the Newman-Penrose frame derivations in 5-dimensions but we have not written out in full all the remaining Newman-Penrose equations (Ricci and Bianchi identities) due to their length. This complete set of equations is much larger than the 4-dimensional set of Newman-Penrose equations but this fact does not render the 5-dimensional Newman-Penrose equations less useful. In fact our expressions may be more suited in computations which seek to find exact solutions because in that case one normally needs to write out all the Ricci and Bianchi identities and, as explained before, the introduction of complex quantities arising from the spin formalism permits us to work with a set of less  quantities and eventually less equations.  
For example, in particular cases in which a number of components of the curvature spinor vanish we may expect significant simplifications just as it happens in the 4-dimensional Newman-Penrose formalism. Other important application in which the 5-dimensional Newman-Penrose formalism developed by us could really show its advantage happens when one wishes to study how to extend a known 4-dimensional exact solution (or group of exact solutions) to five dimensions. For example, we may be interested in studying all the possible extensions of the four dimensional Petrov type D vacuum solutions to five dimensions. In this particular case one knows from the 4-dimensional analysis that a spin tetrad such that the conditions $\Psi_0=\Psi_1=\Psi_3=\Psi_4=0$ can be chosen. Something similar would happen if we are looking for extensions of 4-dimensional Petrov type N solutions to five dimensions (in this case the conditions are $\Psi_1=\Psi_2=\Psi_3=\Psi_4=0$).

Another interesting property is the fact that some of the quantities used in the extended Newman-Penrose formalism behave as {\em weighted quantities} under the transformation  
\begin{equation}
o^A\rightarrow H o^A\;,\quad
\iota^A\rightarrow\frac{1}{H}\iota^A\;,\quad
\tilde{o}^A\rightarrow \overline{H} \tilde{o}^A\;,\quad
\tilde{\iota}^A\rightarrow \frac{1}{\overline{H}}\tilde{\iota}^A, 
\label{eq:spin-boost-transformation}
\end{equation}
where $H$ is a complex parameter. This transformation keeps the symplectic metric $\epsilon_{AB}$ and hence it can be related to a Lorentz transformation of the metric $g_{ab}$. If under (\ref{eq:spin-boost-transformation}) a scalar $Z$ changes according to the rule
$$
Z\rightarrow H^p\overline{H}^q Z,
$$ 
then it is said that the scalar $Z$ is a $(p,q)$-weighted quantity. 
One can then define the boost weight and the spin weight of $Z$ in a similar fashion as in four dimensions 
$$
\mbox{boost weight:}\ \frac{p+q}{2}\;,\ \mbox{spin weight:}\ \frac{p-q}{2}. 
$$ 
All the components of the curvature spinors defined in (\ref{eq:weyl-spinor-complex-components})-(\ref{eq:ricci-spinor-real-components}) are $(p,q)$-weighted quantities for certain integers $p$, $q$. Also most of the spin coefficients are weighted quantities and indeed only $\alpha$, $\beta$, $\gamma$, $\epsilon$, $\theta$ are non-weighted \cite{AGPLODE}.
This raises the possibility of an extension of the Geroch, Held and Penrose formalism 
\cite{GHP} to dimension five. In fact it is not very difficult to introduce the weighted differential operators in terms of the Newman-Penrose frame differentials. One concludes that the weighted differential operators constructed from $D$, $\Delta$, $\delta$ and $\bar\delta$ coincide with the 4-dimensional definitions of, respectively, \TH, \TH$'$, \dh, \dh$'$ as shown in \cite{AGPLODE}. 

Other interesting issue is the algebraic classification of the Weyl spinor. This has been tackled in \cite{SMET1} where an invariant classification of this spinor was put forward. Under this classification there are twelve different ``Petrov types'' of 
the Weyl spinor so it would be interesting to find out how one can characterise these Petrov types in terms of conditions involving the components of the Weyl spinor (some cases are already analysed in \cite{SMET1}). Alternatively, one could try to apply the alignment theory  directly to the Weyl spinor and devise a classification for it as in \cite{ALIGNED-THEORY,COLEY-REVIEW}. This theory is based on studying the boost weights of those scalar components of the Weyl tensor which do not vanish on a suitably chosen frame and hence it is clear that we could follow the same procedure if we used the scalar components of the Weyl spinor and the notion of boost weight discussed above. Indeed some Petrov types adopt a simpler form when we work with the components of the Weyl spinor. For example, a spacetime is of Petrov type D if and only if the components of the Weyl spinor different from zero are those of boost weight zero. These are
$$
\ \Psi_2\;,\ \Psi_{11}\;,\ \ ^*\Psi_2\;,\ \Psi^*_2\;,\ \Psi_{02}\;.
$$     
One could take this as the starting point of a systematic study of all the possible 5-dimensional (vacuum) type D exact solutions of the Einstein equations. To that end one sets to zero in the 5-dimensional Newman-Penrose equations all the curvature scalars except those shown in previous equation (if we do not work in vacuum then we need to retain the components of the Ricci spinor)
and then checks the consistency with the commutation relations shown below. Work in this direction has been already started in \cite{AGPLODE} for the vacuum case using the extension of the GHP formalism mentioned above .

The 5-dimensional spinor calculus is now being implemented in the {\em Mathematica} package {\em Spinors} \cite{Spinors}, which is part of the {\em xAct} system \cite{xAct}.

\section*{Acknowledgements}
AGP is supported by a postdoctoral contract by Ghent university. 
JMM was supported by the French ANR Grant BLAN07-1\_201699 entitled
``LISA Science'', and also in part by the Spanish MICINN Project
2008-06078-C03-03. We thank Dr. Lode Wylleman for pointing out some typos.

\appendix

\section{Symplectic metrics on a vector space}
\label{appendix:symplectic}
Let ${\mathbf V}$ and ${\mathbf V}^*$ be, respectively, a vector space (real or complex) and its dual and
let us use small Latin characters $a,b,c,\dots $ to denote the abstract indices 
of the elements of the tensor algebra built with $\mathbf V$ and $\mathbf V^*$, which is $\mathfrak{T}({\mathbf V})$. 
We introduce next two quantities $M_{ab}$ and $T^{ab}$ establishing linear isomorphisms $M:{\mathbf V}\rightarrow {\mathbf V}^*$ and 
$T:{\mathbf V}^*\rightarrow {\mathbf V}$ in the following way
\begin{equation}
v_a \equiv M_{ab} v^b
\qquad {\rm and} \qquad
\omega^a \equiv T^{ab} \omega_b \qquad {\rm for\ any}\ v^a\in {\mathbf V},\ \omega_a\in {\mathbf V}^*.
\label{up-down}
\end{equation}
Note the convention of having only the second indices
of $M$ and $T$ as contracted indices. Previous isomorphisms are generalised to 
$\mathfrak{T}(\mathbf V)$ in the obvious way and shall be referred to as the operation of ``raising and lowering of indices''. In addition we impose that 
$T=M^{-1}$, and so (\ref{up-down}) implies
\begin{equation}
T^{ab} \, M_{bc} = \Delta^a{}_c ,
\end{equation}
with $\Delta^a{}_c$ the identity on ${\mathbf V}$ (Kronecker delta on ${\mathbf V}$), and
\begin{equation}
M_{ab} \, T^{bc} = \delta_a{}^c ,
\end{equation}
with $\delta_a{}^c$ the identity on ${\mathbf V^*}$ (Kronecker delta on 
${\mathbf V^*}$).

We can change indices with the Kronecker delta tensors, and now we can also
raise and lower indices making use of the $M$ and $T$ isomorphisms. Suppose now that we wish to compute the product $M_{ab} \Delta^b{}_c$. We can either
lower an index of $\Delta$ or change an index of $M$. We conclude:

\begin{equation}
M_{ac} = \Delta_{ac},
\label{m-d}
\end{equation}
and similarly
\begin{equation}
T^{ac} = \delta^{ac} .
\end{equation}
We can also see that
\begin{equation}
T^{ab} = M^{ba} \qquad {\rm and} \qquad
T_{ab} = M_{ba} ,
\end{equation}
independently of the symmetries of $M$ and $T$, which could even have
no symmetry at all.
Concluding, we always have, for indices of any character, and any symmetry:
\begin{equation}
T^{ab} = M^{ba} = \Delta^{ba} = \delta^{ab}\;,\quad
T_{ab}=M_{ba}=\Delta_{ba}=\delta_{ab}\;,\quad
T_a^{\phantom{a}b}=\delta_a^{\phantom{a}b}\;,\quad
M^a_{\phantom{a}b}=\Delta^a_{\phantom{a}b}.
\label{eq:up-eq}
\end{equation}
The four quantities $T$, $M$, $\Delta$, $\delta$ are essentially the same. Let us take, for clarity, only $T$. It always obeys:

\begin{equation}
v^a=T^{ab} v_b , \qquad
v_a=T_a{}^b v_b , \qquad
v_b=v^a T_{ab} , \qquad
v^b=v^a T_a{}^b.
\end{equation}

\noindent However, the following are generically undefined

\begin{eqnarray}
T_{ab} v^b , \qquad
T^a{}_b v^b, \qquad
v_a T^{ab}  , \qquad
v_a T^a{}_b,
\end{eqnarray}
unless $T^{ab}$ has a definite symmetry which means that either $T^{ab}$ is symmetric or antisymmetric. When this is the case we deduce from (\ref{eq:up-eq}) that $T_{ab}$ $M^{ab}$, $M_{ab}$, $\Delta_{ab}$, $\delta_{ab}$, 
$\Delta^{ab}$ and $\delta^{ab}$ all inherit the symmetry of $T_{ab}$ and indeed we could just regard the quantity $T^{ab}$ as fundamental and the remaining ones as derived from it, keeping the symbol $T$ as the kernel letter for all of them. 
Also using (\ref{eq:up-eq}) one may deduce
\begin{equation}
\delta_{a}^{\phantom{a}b}=\Delta^{b}_{\phantom{b}a},
\label{eq:delta-delta}
\end{equation}
if $T^{ab}$ is symmetric and 
$$
\delta_{a}^{\phantom{a}b}=-\Delta^{b}_{\phantom{b}a},
$$
if $T^{ab}$ is antisymmetric. In the case of $T^{ab}$ being symmetric then 
one introduces a quantity $\delta^a_b$ to mean either $\delta_{a}^{\phantom{a}b}$ or 
$\Delta^{b}_{\phantom{b}a}$ and no confusion can arise. However, if $T^{ab}$ is antisymmetric and we insist on keeping only one delta symbol $\delta^a_b$ we need to specify also whether $\delta^a_b$ refers to $\delta_{a}^{\phantom{a}b}$ or to $\Delta^{b}_{\phantom{b}a}$. We believe that to keep the notation $\delta^a_b$ in this context is somewhat confusing and one should instead pick up one of the ``deltas'' as the fundamental one and regard the other as a derived quantity. For example if we agree to take $\delta_a^{\phantom{a}b}$ as the fundamental quantity (as we do in our discussion in section \ref{sec:spin-5d}) then we have 
$$
\delta^a_{\phantom{a}b}=\Delta^a_{\phantom{a}b}=-\delta_b^{\phantom{b}a},
$$  
and no confusion arises.

\end{document}